\documentclass[reprint,
nofootinbib,
 amsmath,amssymb,
 aps,
]{revtex4-2}

\usepackage{graphicx}% Include figure files
\usepackage{dcolumn}% Align table columns on decimal point
\usepackage{bm}% bold math
\usepackage{braket}
\usepackage{xcolor}
\usepackage{footnote}
\usepackage[normalem]{ulem}
\usepackage{mathtools}
\usepackage[makeroom]{cancel}
\newcommand\underrel[3][]{\mathrel{\mathop{#3}\limits_{%
      \ifx c#1\relax\mathclap{#2}\else#2\fi}}}

\begin{document}

\title{Quantum information engines: Bounds on performance metrics by measurement time}% Force line breaks with \\

\author{Henning Kirchberg}
\email{henning.kirchberg@chalmers.se}
\affiliation{
Chalmers University of Technology, Department of Microtechnology and Nanoscience - MC2, Göteborg, Sweden, 41296
}
\author{Abraham Nitzan}
\affiliation{
 University of Pennsylvania, Department of Chemistry, Philadelphia, PA, U.S., 19104
}

\date{\today}% It is always \today, today,

\begin{abstract}
Information engines, sometimes referred to as Maxwell Demon engines, utilize information obtained through measurement to control the conversion of energy into useful work. Discussions around such devices often assume the measurement step to be instantaneous, assessing its cost by Landauer's information erasure within the measurement device. While this simplified perspective is sufficient for classical feedback-controlled engines, for nanoengines that often operate in the quantum realm, the overall performance may be significantly affected by the measurement duration (which may be comparable to the engine's cycle time) and cost (energy needed to create the system-meter correlation).
In this study, we employ a generalized von-Neumann measurement model to highlight that obtaining a finite amount of information requires a finite measurement time and incurs an energetic cost. We investigate the crucial role of these factors in determining the engine's performance, particularly in terms of efficiency and power output. Furthermore, for the information engine model under consideration, we establish a precise relationship between the acquired information in the measurement process and the maximum energy extractable through the measurement. We also discuss ways to extend our considerations using these concepts, such as in measurement-enhanced photochemical reactions.
\end{abstract}

\maketitle

%\tableofcontents
\section{Introduction} A prominent example of energy conversion devices are heat engines which operate between reservoirs at different temperatures. Alternatively, a single heat bath may be used as the energy source in feedback-controlled devices \cite{sag2012,vid2016,cot2017,Cottet2018,elo2017,elo2018,mon2020,bre2021}, referred to below as information engines (IEs), in which information about the system's state, obtained by some "Maxwell demon" - a general intelligent outside controller - is used to guide the engine's operation \cite{MaxwellBook,ZurBook,mar2009,man2012,def2013,str2013,hor2013,bar2013}. In general, the second law in these engines is accounted for by the increase in entropy during the demon's restoration to its initial state, also implying a minimal added operation cost \textemdash  Landauer's erasure work \cite{lan1961}. In the case of a classical Maxwell’s demon, the measurement is ideally arbitrarily precise, and system and demon are classically correlated.
In the fully quantum version of such devices, the demon's acquisition of information is often described as a quantum measurement process. As in the case of a classical meter, a quantum (von-Neumann) measurement involves an interaction between the system (S) being measured and a quantum meter (M), which leads to a correlated system-meter state \cite{jun2024,VonNeumannBook}, so that a subsequent observation of the meter yields information about the system. Such quantum measurement models that were the subjects of recent theoretical studies \cite{gur2020,tar2023} pose several practical and conceptual difficulties. First, the aforementioned studies have demonstrated that achieving ideal quantum measurements, which correlate the state of the meter precisely with the state of the system, is unfeasible with finite resources (finite energy, finite time, and finite complexity, i.e., dimensionality of the meter space). Consequently, real measurements are inherently nonideal owing to the limitations of finite resources and their dependence on the amounts of resources allocation determines the efficiency and operating power of the associated energy conversion device. Second, the state of a quantum meter can only be determined by a quantum measurement, leading to the well-known conundrum of a sequence of subsequent measurements that need to be truncated by some supplementary assumption to go from the quantum to the classical 'objective' state description of the meter \cite{MenskyBook}.
An alternative technique to describe real and nonideal measurements with possible measurement errors is to coarse grain over the meters degrees of freedom, leading to positive operator-valued measures (POVM) like Kraus operators that act on the system in some assumed form [see, e.g. \cite{jac2006,ann2022,och2018,wis1994}]. Such an approach makes it possible to investigate important thermodynamic characteristics of information engines (such as the aforementioned Landauer lower bound on the unavoidable dissipation, which recent studies have shown to be compatible with fluctuation theorems of stochastic thermodynamics \cite{sag2010,toy2010,sag2012}). However, in this framework, the actual dynamics of the coupled system and meter is not explicitly described, making it impossible to address the time and related cost involved in acquiring information through measurement.

The energetic and temporal aspects of measurement-driven information engines were subjects of several recent studies \cite{cot2017,elo2018,bre2021}. A notable experimental example involves measuring a qubit state in a microwave cavity where the cavity acts as the measuring entity, or "demon" \cite{cot2017}. In this work the experimental setup is an autonomous driven process in which an internal process (photon occupation of the cavity identified as demonic "measure") affects another internal process (energy extraction by resonant-stimulated emission) making the identification of cost and gain open to interpretation. 

Generally, information gain is not instantaneous; rather, it is characterized by the correlation established between the system and the measurement apparatus, quantified by the mutual information acquired over time and the (energy) cost of coupling and decoupling the quantum system and the meter in order to affect their mutual correlation. Consequently, measurement time and cost are intrinsically linked and must be considered when evaluating performance metrics such as efficiency and operational power in cyclic IEs. This interrelation between the information gained by quantum measurement and the measurement time and energy cost has not been extensively explored in the literature, although there are some significant early exceptions (see \cite{ghi1979,bra2013,bus2010,bre2021}). This paper examines for the first time this interrelation and its consequences for performance metrics for a cyclic operating IE.

As already alluded to above, a crucial step in the measurement process is the "objectification" of the measurement outcome \cite{kor2017}, which describes the transition of the apparatus from quantum to classical behavior, thereby transforming the measurement outcome into an objective fact that can be verified by independent observers. The outcome of this step provides the information that drives the IE. One might argue that the quantum-to-classical transition is an inconsequential issue since, in practice, measurements typically involve large measurement devices. However, as we develop smaller measurement devices down to the nanoscale, this question becomes increasingly important \cite{tar2023}. This is particularly true if the characteristic operational time scale of the device is of the same order as the time required to acquire information about it. Although this paper does not solve the "measurement problem" at the quantum-classical interface \cite{MenskyBook,tar2023}, we propose a different route that circumvents this problem: We place the step of quantum-to-classical transition, the Heisenberg cut \cite{MenskyBook,han2022}, one step further away from the physical system, that is, between a quantum meter and a classical meter. This makes it possible to fully consider the joint evolution of the coupled quantum system and meter to explore the role played by the measurement duration (i.e., the time during which the system and meter are coupled) and the energetic cost required for information acquisition.

This paper is structured as follows. First, we outline the general setup of an IE engine cycle in Section \ref{IEmodel}, where information about a working system, acquired through time-dependent interaction with a meter, is used to extract useful work. Section \ref{example} details a specific implementation of the IE using a two-level system monitored by a free quantum particle. In Section \ref{operation}, we examine the operation of this specific IE implementation and evaluate its performance in power and efficiency. Finally, Section \ref{conclusion} provides a conclusion for this paper.

%%%%%%%%%%%%%%%%%%%%%%%%%%%%%%%%%%%%%%%%%%

\section{A general IE model}
\label{IEmodel}

\begin{figure}[h!!!!!]
\centering
\includegraphics[width=\linewidth]{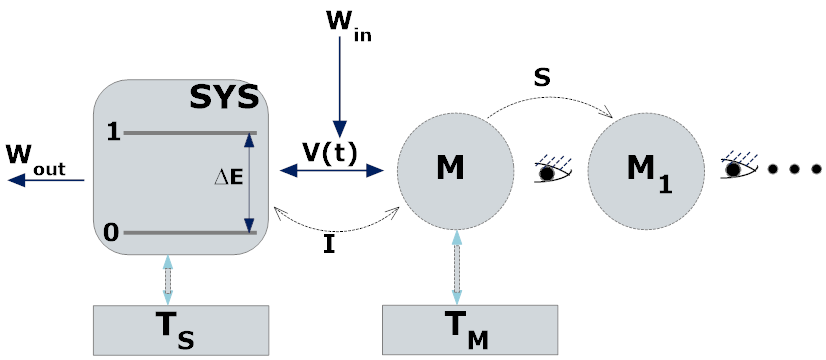}
\caption{\label{fig1A} General schematics of an IE model. A system (S) and a meter (M), each coupled to their own thermal bath of temperatures $T_{\rm S}$ and $T_{\rm M}$, respectively, are entangled by an time-dependent interaction $V(t)$. The state of M is projectively monitored by a classical meter M$_1$ accompanied by an entropy flow $S$ between M and M$_1$. Information $I$ on the state of the system, obtained by the meter, is used to extract energy $W_{\rm out}$ form the system bath. The measurement time and its energy cost $W_{\rm in}$ are computed and used to calculate the energy efficiency and operating power.}
\end{figure}
A general scheme for an IE based on a quantum von-Neumann measurement is shown in Fig.\ \ref{fig1A} and is characterized by the following steps:

(i) \textit{Initialization:} The system (S) and meter (M) are initially in thermal equilibrium with their respective bath. In the energy range of interest the system is assumed to have a discrete energy described by the Hamiltonian $\hat{H}_{ \rm S}= \sum_i E_i \ket{i}\bra{i}$. Its initial state is thus represented by a density operator $\hat{\rho}_{\rm S,in}=\sum_i p_i \ket{i}\bra{i}$, $p_i=\exp{(-\beta_{ \rm S} E_i)}/Z_{\rm S}$ being the thermal population of the state of the system with the partition function $Z_{ \rm S}=\mathrm{tr}[\exp{(-\beta_{ \rm S} \hat{H}_{ \rm S})}]$ and $\beta_{ \rm S}=k_B T_{\rm S}$. In the example considered below we specify to a two-level system. The meter is in a thermal state of its own, characterized by a temperature $T_{\rm S}$, so that $\hat{\rho}_{\rm M,in}=\exp{(-\beta_M \hat{H}_{\rm M})}/Z_{ \rm M}$ with $Z_{ \rm M}=\mathrm{tr}[\exp{-(\beta_{ \rm M} \hat{H}_{ \rm M})}]$ where $\hat{H}_{ \rm M}$ being the Hamiltonian of the meter \cite{footnote18}.

(ii) \textit{Unitary evolution:} After decoupling the system (S) and the meter (M) from their respective baths, the measurement process occurs by coupling S with the meter M with each other which is characterized by a coupling operator $\hat{V}(t)$ that is different from zero during the measurement interval $0 \leq t \leq t_m$. This measurement is designed to determine the energy state of the system and is therefore taken in the form  $\hat{V}(t) = \sum_i g_i(t) \ket{i}\bra{i} \otimes \hat{M}$, where $\ket{i}$ are eigenstates of $H_S$, $\hat{M}$ is an operator in the meter subspace and $g_i(t) \neq 0$ during the measurement interval $0 < t < t_m$. The condition $[\hat{M}, \hat{H}_{\rm M}] \neq 0$ ensures that the states of the system and the meter become correlated.
Given the instantaneous switching on at $t=0^+$ and switching off at $t=t_m^-$ of the system-meter interaction, the change of energy can be associated to cost of this switching process, the measurement cost (see detailed discussion in Sec.\ IV in the Supplemental Material \cite{SM1}),
\begin{align}
\label{intrE1}
W_{\rm meas}(t_m) &\equiv \textrm{tr} [\big(\hat{\rho}(t_m)-\hat{\rho}(0)\big)\big\{\hat{H}_{\rm S}+\hat{H}_{ \rm M}\big\}] \\ \notag
&= \textrm{tr} [\big(\hat{\rho}(t_m)-\hat{\rho}(0)\big)\hat{H}_{ \rm M}],
\end{align}
while the last line in Eq.\ \eqref{intrE1} holds since $[\hat{H}_{ \rm S},\hat{H}(t)] = 0$, i.e., the energy invested in the measurement ends up in the meter. In Eq.\ \eqref{intrE1} we have the joint density matrix of the system-meter evolution $\hat{\rho}(t)= \hat{U}(t) [\hat{\rho}_{S,in}\otimes \hat{\rho}_{M,in}] \hat{U}^\dagger (t) $ with $\hat{U}(t)=\mathcal{T}\exp{\{-\frac{i}{\hbar}\int_0^{t}dt'\hat{H}(t')\}}$ where $\mathcal{T}$ is the time ordering operator and $\hat{H}(t)=\hat{H}_{\rm S}+\hat{H}_{ \rm M}+\hat{V}(t)$, while $\hat{\rho}(0)=\hat{\rho}_{\rm S,in}\otimes \hat{\rho}_{\rm M,in}$.

(iii) \textit{Projective measurement \& information gain:} After time $t_m$, systems S and M become decoupled from each other. The state $\ket{m}$ (an eigenstate of $\hat{H}_{ \rm M}$) of the meter is then determined by an associated projective measurement by a classical meter M$_1$ that yields the reduced state after measuring $\hat{P}(m,t_m)\equiv \braket{m|\hat{\rho}(t_m)|m}$, $\hat{\rho}(t_m)$ being the joint system-meter density operator. The result of this measurement provides the necessary information used to drive the IE. Since M and M$_1$ are classically correlated, this correlation incurs no cost beyond the classical Landauer cost associated with information theory, given by $W_{\rm L} = T_{\rm M1} S$, where $T_{\rm M1}$ is the temperature of meter M$_1$ and $S$ is the entropy flow between M and M$_1$ \cite{lan1961}. In our analysis, we assume that the classical readout reservoir has a temperature of zero, $T_{\rm M1}=0$, allowing us to disregard the Landauer erasure work. Given that the temperature of the classical meter can be chosen somewhat arbitrarily, setting $T_{\rm M1}=0$ is a justifiable approach to explore potential maximal energy conversion processes in feedback-controlled processes through measurement, as discussed in \cite{elo2017,jor2020}. By shifting the projective measurement or Heisenberg cut - the interface between quantum events and a classical observer's information \cite{atm1997} - one step further from the physical system \cite{MenskyBook,gur2020,lat2024} allows one to for a separate analysis of the entangling evolution of the coupled system and meter, the duration of the measurement (i.e., the time during which the system and meter are coupled), and its actual energetic cost $W_{\rm in}=W_{\rm meas}+W_{\rm L}$. With the reasoning as discussed above, we take $W_{\rm L}=0$, such that  $W_{\rm in}\equiv W_{\rm meas}$.

(iv) \textit{Work extraction:} The acquired knowledge about the system is utilized to convert thermal energy (at the system temperature $T_{\rm S}$) into useful work. We will discuss different models for estimating the extracted work in the next section.

(v) \textit{Restoration:} The IE cycle concludes by returning both the system and the meter to their initial states, achieved by coupling them to their respective thermal baths and resetting the classical meter.

The ratio of the net work extracted (work extraction minus measurement energy cost) to the energy invested along the working cycle determines the device's efficiency, while the net work extracted per cycle time represents its power. Note that the measurement time, $t_m$, serves as a lower bound for the cycle time, which means that this calculation will provide an upper bound on the operational power. In Section \ref{operation}, we explore these quantities within a specific device model.

The IE cycle described above is general, and specific IE models will differ by their realization of the system and meter and the associated input and output energies. One such specific example is described and analyzed in the next Section  \ref{example}. 

\section{Two-level system monitored by free-particle meter}
\label{example}
The working entity S (Fig.\ \ref{fig1A}) is taken to be a two-level system (TLS) with energy levels $E_0=0$ and $E_1=\Delta E>0$ for the lower ($\ket{0}$) and upper state ($\ket{1}$), respectively. This system is monitored by coupling it to a meter M modeled as an otherwise free particle. The IE operation cycle starts with the system and meter at thermal equilibrium with their respective baths, but during the measurement (while mutually interacting and when the meter state is determined to acquire information about the system) they are assumed to be decoupled from these baths. The Hamiltonian of this combined TLS-M-system reads $\hat{H}=\Delta E \ket{1}\bra{1}  +\frac{\hat{p}^2}{2} + \hat {V} (t)$ where $\hat{p}$ is the mass weighted momentum operator of the meter and $\hat {V} (t)$ is the system-meter interaction. In the present analysis we assume a sudden switch on and off of the interaction to a constant value $g$ and take it to be 
\begin{align}
\label{coupling2}
 \hat{V}(t)=gD(t)\cdot \hat{x}\otimes\ket{1}\bra{1} = D(t)\hat{V},
\end{align}
 where $\hat{x}$ is the meter position operator and $D(t)=1$ for $0\leq t \leq t_m$, while $D(t)=0$ otherwise. This form implies that during the measurement interval the meter responds to the system only if the latter is in state 1 and that its response is expressed by a momentum shift (an often used model \cite{art1965} which is potentially realizable in practice \cite{mac2013}).
The detailed IE engine cycle along the steps described in the previous Section \ref{IEmodel} are:

(i) \textit{Initialization:} 
The initial density matrix of system and meter is defined as 
\begin{align}
&\hat{\rho}(t=0)=\hat{\rho}_S(t=0)\otimes \hat{\rho}_M(t=0),
\end{align}
with
\begin{align}
\label{in1} 
\hat{\rho}_S(t=0)= a\ket{0}\bra{0}+b\ket{1}\bra{1}, 
\end{align}
and \begin{align}
\label{meterIn}
\hat{\rho}_M(t=0)=\bigg( \frac{1}{2 \pi \hbar^2 k_B T_{ \rm M}} \bigg)^{1/2} e^{-p^2/2 k_B T_{ \rm M} } \ket{p}\bra{p}.
\end{align}

In Eq.\ \eqref{in1} $a$ and $b$ are real positive numbers satisfying $T_{ \rm S}=\Delta E/k_B [\ln(a/b)]^{-1}$ and $a+b=1$. Similarly, Eq.\ \eqref{meterIn}, represents the meter at thermal equilibrium at temperature $T_{ \rm M}$, written in the momentum representation.

(ii) \textit{Unitary evolution:} Uncoupled from their respective thermal baths, the system and meter evolve under the Hamiltonian $\hat H$ during the time interval $(0,t_m)$, leading to an entangled state described by the density matrix $\hat{\rho}(t_m)=e^{-i \mathcal{T}\int_0^{t_m} dt'\hat{H}(t')/\hbar} \hat{\rho}(0) e^{i \mathcal{T}\int_0^{t_m} dt'\hat{H}(t')}/\hbar$. The energy needed to create the system-meter entanglement, see Eq.\ \eqref{intrE1} and the detailed derivation in Sec.\ IV of the Supplementary Material \cite{SM1}, is given by
\begin{align}
\label{intrE}
W_{\rm meas}(t_m) \equiv \textrm{tr} [\hat{\rho}(0) \hat{V}]-\textrm{tr}  [\hat{\rho}(t_m) \hat{V}] = \frac{bgt_m^2}{2} ,
\end{align}
where $\textrm{tr} [\dots]\equiv \int dp \sum_{i=0,1} \bra{p} \bra{i} \dots \ket{i}\ket{p}$. Eq.\ \eqref{intrE} follows from Eq.\ \eqref{intrE1} by using the fact that the switching is instantaneous to a constant value. For our choice of initial states and system-meter interaction $\textrm{tr} [\hat{\rho}(0) \hat{V}]=0$, namely switching on the interaction costs no energy. 

(iii) \textit{Projective measurement \& information gain:} Following the unitary evolution, the state of the meter is projectively determined in the basis of eigenstates of the momentum operator. It is assumed that this process is instantaneously registered in the classical meter M$_1$. As discussed above, we assume that this step does not incur an additional energy cost. The conditional probability of the TLS to be in state $i=0;1$ given the meter outcome $p$ is thus determined by 
\begin{align}
    P_i( t_m|p)= \frac{\bra{i} \braket{p|\hat{\rho}(t_m)|p}\ket{i}}{Q(p,t_m)} = \frac{P_i(p,t_m)}{Q(p,t_m)},
\end{align}
where $Q(p,t_m)=\sum_{i=0}^1 P_i(p,t_m)$, while for our IE model \cite{SM1},
\begin{align}
P_0(p, t)&= \sqrt{\frac{1}{2 \pi  k_B T_{\rm M}}}   a e^{-\frac{p^2}{2 k_B T_{\rm M}}}, \\
P_1(p, t)&= \sqrt{\frac{1}{2 \pi k_B T_{\rm M}}} b e^{-\frac{(p+g t)^2}{2 k_B T_{ \rm M}}}.
\end{align} 

Next, define the conditional density matrix $\hat{P}(t_m|p)$ to be in state $i=0$ ; $1$ given  the meter outcome $p$, which reads
\begin{align}
\label{meterout}
    \hat{P}(t_m|p)=P_0(t_m|p) \ket{0}\bra{0}+ P_1(t_m|p) \ket{1}\bra{1}.
\end{align}

The information gain, $I(t_m)$, in this measurement process can be quantified by averaging the conditional system entropy
$S(t_m|p)=-k_B \sum_{i=0}^{1}  P_i(t_m|p) \ln{P_i(t_m|p)}$ over an ensemble of identical measurements by
\begin{align}
S(t_m) &= \int dp Q(p,t_m) S(t_m|p) \\ \notag &= -k_B \int_{-\infty}^\infty dp \sum_{i=0}^{1}  P_i(p,t_m) \ln{P_i(t_m|p)}
\end{align}
leading to \cite{footnote4,sha1948} 
\begin{align}
\label{know}
I(t_m)\equiv S(0)-S(t_m),
\end{align}
which is equal to the mutual information expression associated with the measurement process, see Sec.\ III in the Supplementary Material \cite{SM1}. 

(iv) \textit{Work extraction:} Given the measurement result $p$, the state of the working system is given by Eq.\ \eqref{meterout}. Consider first the ergotropy of this state, namely the maximum work that can be extracted from it under a unitary transformation \cite{all2004}
\begin{align}
    \label{ergotropy}
    &W_{\rm erg}(t_m|p) \equiv \textrm{tr}_{\rm S} [\hat{P}(t_m|p) \hat{H}_{\rm S}] -\min_{\hat{U}}\textrm{tr}_{\rm S}  [\hat{U}\hat{P}(t_m|p)\hat{U}^\dagger \hat{H}_{\rm S}]\\ \notag
    &=\Delta E(P_1(t_m|p)-P_0(t_m|p))\Theta(P_1(t_m|p)-P_0(t_m|p)),
 \end{align}
where $\Theta(x)$ is the heavy side function $\Theta(x)=1$ for $x>0$ and $\Theta(x)=0$ otherwise. 

Although this work \eqref{ergotropy} is defined as an abstract concept, it is of interest to demonstrate a potentially practical implementation of extracting this energy: A $\pi$-pulse can be used to interconvert between the molecular ground and excited states with probability 1. To obtain a net gain from applying such a pulse, the molecular population needs to be inverted. Define $p'$
\begin{align}
    P_0(t_m|p')=P_1(t_m|p')=0.5,
\end{align}
where we note that $W_{\rm erg}(p\geq p')=0$. This implies that sending a $\pi$-pulse photon onto the system results in net loss if $p>p'$ and net gain if $p<p'$. This gain through state-inversion $\Delta E [P_1(t_m|p)-P_0(t_m|p)]$ is equal to the ergotropy.
Averaging over the measurement results leads to
\begin{align}
\label{ergotropy2}
W_{\rm erg}(t_m)=\Delta E \int_{-\infty}^{p'} Q(p,t_m) [P_1(t_m|p)-P_0(t_m|p)].
\end{align}
Note that using Eq.\ \eqref{ergotropy2} as a quantifier for the measurement enhanced gain is based on the assumption that preparing the $\pi$-pulse costs only the energy embedded in the pulse itself, and spontaneously emitted photons do not contribute to the gain.

We emphasize that while the general definition of ergotropy does not explicitly consider the time needed to execute the optimal unitary transformation, the demonstration that this optimal extraction can only be done with a photon $\pi$-pulse indicated that this part of the IE cycle can be carried out on a timescale of order $\hbar/\Delta E$ ($10^{-12}$-$10^{-15}$s in molecular systems).

(v) \textit{Restoration:} Following the measurement-driven extraction of useful energy, the engine cycle is closed by restoring the TLS and meter to their initial thermal states. The temperature of the final state of the system is given by (see discussion in Sec.\ V in the Supplementary Material \cite{SM1})
\begin{align}
\label{temperature}
    T_{ \rm p}&(p,t_m)=\frac{\Delta E}{k_B} \bigg(\ln{\bigg[\frac{P_1(t_m|p)}{P_0(t_m|p)}\bigg]}\bigg)^{-1} \\ \notag &\times \Theta(P_1(t_m|p)-P_0(t_m|p))\\ \notag
    & + \frac{\Delta E}{k_B} \bigg(\ln{\bigg[\frac{P_0(t_m|p)}{P_1(t_m|p)}\bigg]}\bigg)^{-1} \Theta(P_0(t_m|p)-P_1(t_m|p)).
\end{align}

The final state of the system at the end of the IE cycle is in thermal equilibrium with the bath $T_{\rm S}$. A common scenario is to achieve this by spontaneous thermal equilibration (which may be fast as $\sim10^{-12}$s for molecular-scale engines) without further gain of useful work. It is interesting, however, to consider also the maximum additional work that can be extracted on way to full thermal relaxation by an adiabatically slow Carnot process. This additional gain, obtained when a TLS at the initial temperature $T_{ \rm p}$ comes to a final equilibrium with a bath at temperature $T_{ \rm S}$ can be obtained by using incremental Carnot steps in which a high-temperature bath releases an amount of heat $d\mathcal{Q}$ on way to equilibration with a low temperature bath.
The maximal part of this heat that can be converted to work is
\begin{align}
dW = d\mathcal{Q} \left(1 - \frac{T_{\text{low}}}{T_{\text{high}}}\right).
\end{align}

At each incremental step, the change in the TLS temperature can be calculated from its know heat capacity
\begin{align}
\label{heatCap}
    C(T)=k_B \beta^2 \frac{d^2\ln{Z_S}}{d\beta^2}=  \frac{\Delta E^2}{k_B T ^2}\frac{e^{-\frac{\Delta E}{k_BT}}}{\big [1+e^{-\frac{\Delta E}{k_BT}}\big]^2},
\end{align}
$Z_S=1+\exp{(-\beta \Delta E)}$ is the partition function of the TLS of specific temperature $T=k_B^{-1}\beta^{-1}$.

Integrating along the relaxation path and averaging over measurement results leads to (see Sec.\ V in the Supplementary Material \cite{SM1})
\begin{align}
\label{thermalwork3}
    W_{\rm th} = \int_{-\infty}^\infty dp Q(p,t_m)\int_{T_{\rm p}(p,t_m)}^{T_{\rm S}} dT C(T) \left(\frac{T_{\rm S}}{T} -1\right),
\end{align}
valid for both $T_{ \rm p}>T_{ \rm S}$ and $T_{ \rm p}<T_{\rm S}$. The fact that work is generated in both cases just reflects the fact that a Carnot engine generates work irrespective of which bath is hotter. The source of the work is the heat released by the hotter bath. Note that while this extra work can be taken into account in the evaluation of the engine efficiency it is not relevant in consideration of power (given the underlying adiabatically slow Carnot process).

For the meter, any extra energy spent on switching on and off the system-meter connection \eqref{meterIn} will be dissipated into the meter reservoir, denoted as $\mathcal{Q}_{ \rm M} \equiv W_{\rm meas}$. While one could theoretically devise a process to recycle this energy back as work, in the subsequent analysis, we will disregard this possibility.

In quantum IE based on molecular systems, work extraction (e.g., by using a $\pi$-pulse as described above) and thermal relaxation can take place on timescales faster or comparable to the time interval between the coupling and decoupling of the system and the measurement apparatus. Thus, the latter can emerge as the predominant timescale for the cyclic process. Implications on performance metrics of such engines will be discussed in Section \ref{operation}.

\section{Operation and Performance}
\label{operation}
In this Section we show, for the IE model described in Section \ref{example}, some examples for how the IE characteristic parameters affect the engine operation and its performance. In particular, we examine the way gaining information is manifested in the resulting performance characteristics.

\subsection{Information gain and energetic cost}
\begin{figure}[h!!!!!]
\centering
\includegraphics[width=1.05\linewidth]{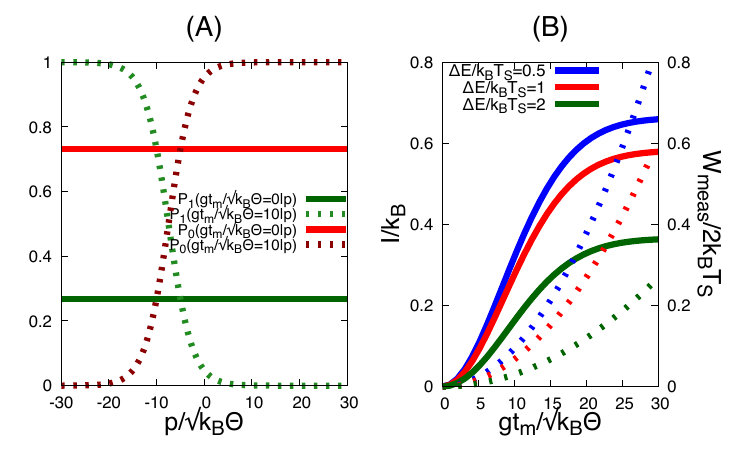}
\caption{\label{fig9} (A) The conditional probability $P_{i=0;1}(t_m|p)$ that a TLS is in state $0$ or $1$ given that the meter outcome is $p$. The TLS is initially in thermal equilibrium at temperature $T_{ \rm S}$. The parameters used are $T_{ \rm S}=300$K, $\Delta E=k_B T_{ \rm S}$ and the initial meter state is given by Eq.\ \eqref{meterIn} with $T_M=300$K. The horizontal red and green line represent the conditional probability $P_0(0|p)=a$ and $P_1(0|p)=b$, respectively. The red and green dotted lines are the conditional probabilities to be in state $1$ and $0$ for $gt_m/\sqrt{k_B\Theta}=10$ where $k_B\Theta=1$meV. (B) The information gain $I(t_m)$ (solid lines, left axis) and the measurement energy cost $W_{\rm meas} (t_m) $ (dotted lines, right axis) plotted against measurement time $t_m$ for different choices of $\Delta E/k_B T_{ \rm S}$ with $T_{ \rm S}=300$K and $T_{ \rm M}=300$K.}
\end{figure}

 Fig.\ \ref{fig9} (A) illustrates the conditional probability $P_{i=0;1}(t_m|p)$ to be in the ground or excited state given the meter outcome $p$. Obviously, $P_{i=0;1}(t_m=0|p) = a, b$ is independent of $p$. For $t_m>0$, the evolution of these probabilities may be written as $a\to a-\delta$ and $b\to b+\delta$, where, if $g$ is chosen positive, $\delta>0$ if the meter outcome is negative ($p<0$), and $\delta <0$ when $p>0$, indicating a higher or lower likelihood that the TLS is in the excited state, respectively. The information gain $I(t_m)$ (Eq.\ \eqref{know}) and measurement cost $W_{\rm meas}(t)$ (Eq.\ \eqref{intrE}) are depicted in Fig.\ \ref{fig9} (B) as function of $t_m$ for different initial TLS states defined by $b/a=\exp{[-\Delta E /k_BT_{ \rm S}]}$. Three observations are notable:

(i) The information gain is a monotonously increasing function of $t_m$ that approaches its maximal value which is the entropy of the initial state, $-k_B(a\ln a +b \ln b)$, as $t_m\to \infty$ (see Sec.\ III in \cite{SM2}). This stands in contrast to the model of Ref.\ \cite{bre2021} where, because of the discrete nature of the meter (another two state system), the dependence on $t_m$ reflects the intrinsic Rabi-oscillation in the system-meter dynamics.

(ii) The rate of information gain, given by the slope $dI(t_m)/dt_m$ of $I(t_m)$ in Fig.\ref{fig9} (B), is maximal near $t_m=b\frac{\sqrt{\langle \delta p(t=0)\rangle^2}}{|d\langle p\rangle/dt|_{t=0}}=\frac{\sqrt{2k_BT_{\rm M}}}{g}$, where $|d\langle p\rangle/dt|_{t=0}$ is the change rate of the expectation value of the momentum of the meter immediately after switching on the interaction between system and meter (see Secs.\ I and II in \cite{SM1}). This characteristic time is determined by the width of the initial meter wavepacket, $\sim k_BT_{ \rm M}$, and the system-meter coupling, $g$. 

(iii) As measurement time $t_m$ increases the information gain $I$ approaches its maximal value. However, in the model considered, the measurement energy cost $W_{\rm meas}=bgt_m^2/2$ (Eq.\ \eqref{intrE}) increases indefinitely, (see dotted lines in Fig.\ \ref{fig9} (B)) resulting in a decreasing trend of the information gain to energy ratio. 

These observations leads to the conclusion that $t_m$ must be finite and constitute a lower bound for the cycle time of the IE as without information no work can be extracted from the system.

\subsection{Work extraction}

\begin{figure}[h!!!!!]
\centering
\includegraphics[width=\linewidth]{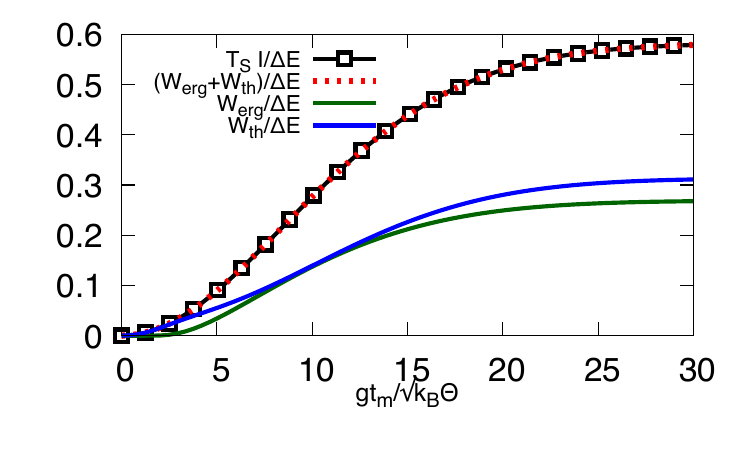}
\caption{\label{fig10} $T_{ \rm S} I$ (Eq.\ \eqref{know}), $W_{\rm erg}+W_{\rm th}$ (Eqs.\ \eqref{ergotropy2}+\eqref{thermalwork3}), $W_{\rm erg}$ (Eq.\ \eqref{ergotropy2}) and $W_{\rm th}$ (Eq.\ \eqref{thermalwork3}) plotted against the measurement time $t_m$ with $T_{ \rm S}=300$K, $\Delta E=k_B T_{ \rm S}$, $k_B\Theta=1$meV, and $T_{ \rm M}=300$K.}
\end{figure}
As seen above, consideration of the maximum work that can be extracted may include the maximum work that can be extracted in the final thermalization of both system and meter. In most practical engines this is not done. For example, we do not usually include the work that can be extracted for a hot car exhaust in evaluating its motor efficiency. We find that it is of interest to consider this contribution for the system as discussed below.

Fig.\ \ref{fig10} shows two components of the measurement provided gain as discussed in Sec.\ \ref{example}: the ergotropy $W_{\rm erg}$, Eq.\ \eqref{ergotropy2}, and the maximum work achievable in the subsequent thermalization, $W_{\rm th}$, Eq.\ \eqref{thermalwork3}, as well as their sum, $W_{\rm tot}=W_{\rm erg}+W_{\rm th}$, evaluated for different values of temperature and plotted against the measurement time $t_m$. The extracted work is seen to increase with measurement time and reach a plateau as $t_m\to \infty$. Two observations are significant:
First, $W_{\rm erg}(t_m\to \infty)= b\Delta E=\frac{\exp{[-\Delta E/k_BT_{ \rm S}]}}{1+\exp{[-\Delta E/k_BT_{\rm S}]}}\Delta E$. Second, the total work that can be extracted from the system ($T_{\rm S}$) thermal bath using the information provided by the measurement is determined by this information according to (see derivation in Sec.\ V in the SM \cite{SM1})
\begin{align}
\label{integral7}
W_{\rm tot}(t_m)=W_{\rm erg}(t_m)+W_{\rm th}(t_m)= T_{\rm S} I(t_m),
\end{align}
where $I(t_m)$ is given by Eq.\ \eqref{know}. We note that the second law of thermodynamics sets the product $T_{\rm S}I$ as an upper bound on the energy that can be extracted in measurement-controlled engines \cite{sag2012,bar2014}. Here we find this relation as an equality, provided that the maximum thermalization work is included in the extracted work.

\begin{figure}[h!!!!!]
\centering
\includegraphics[width=\linewidth]{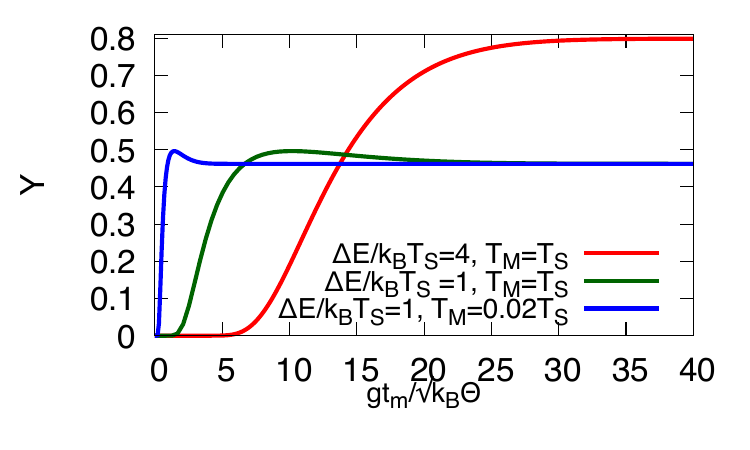}
\caption{\label{fig11} Ratio  $Y$, Eq.\ \eqref{yield}, plotted against the measurement time $t_m$ with $T_{ \rm S}=300$K and $k_B\Theta=1$meV.}
\end{figure}

To further examine the relationship, we define the ratio between the measurement determined ergotropy and the maximum given by Eq.\ \eqref{integral7}
\begin{equation}
\label{yield}
    Y(t_m) \equiv \frac{W_{\rm erg}(t_m)}{T_{ \rm S} I(t_m)}.
\end{equation}
This ration is plotted against $t_m$ for various system temperatures in Fig.\ \ref{fig11}. It is seen to increase with measurement before reaching a plateau, as more energy can be extracted per gained information. The initial rate of gain in $Y$ increases with the accuracy of the meter as expressed by its equilibrium temperature (the measurement is more accurate when the Gaussian peak in Eq.\ \eqref{meterIn} is narrower, namely when the meter is cooler). The small maximum in $Y$ seen for some system parameters (green and blue lines in Fig.\ \ref{fig11}) indicates the possibility of optimal performance in terms of work extraction per information gained. In all scenarios, $Y$ goes to zero when $t_m\to 0$, highlighting the fact that a finite measurement time is needed for the extraction of work assisted by measurement. 

\subsection{Efficiency and Power output}
We define the efficiency of work extraction for information engines (IEs) following \cite{elo2017,elo2018} by
\begin{align}
\label{efficiency}
 \eta (t_m) =\frac{W_{\rm out}(t_m)-W_{\rm meas}(t_m)}{\mathcal{Q}_{ \rm S}(t_m)+W_{\rm meas}(t_m)}.
\end{align}
Here, $W_{\rm meas}$ (see Fig.\ \ref{IEmodel}) is the measurement cost, Eq.\ \eqref{intrE}, while $W_{\rm out}$ is the useful energy gained during the cycle. In the result for engine efficiency displayed in Fig.\ \ref{figIJ} we take it as either $W_{erg}$ (Eq.\ \eqref{ergotropy2}) or $W_{\rm tot}$ (Eq.\ \eqref{integral7}). Note that $\mathcal{Q}_{\rm S}(t_m)$ represents the average heat per cycle extracted from the system's thermal bath to restore the two-level system to its thermal state after work (photon) extraction. This average heat is equivalent to the average work extracted per cycle, denoted as $\mathcal{Q}_{\rm S}(t_m) \equiv W_{\text{out}}(t_m)$. 
For the consideration of power $W_{tot}$ is irrelevant as discussed in Sec.\ \ref{example} (although one may envision scenarios in which part of the thermalization work can be extracted on relevant timescales and hence modify the power) and only $W_{\rm out}=W_{\rm erg}$ is considered in finding bounds on the engine power (denoted $\Pi$). 
\begin{align}
\label{power}
\Pi(t_m) &\leq \frac{W_{\rm erg}(t_m)-W_{\rm meas}(t_m)}{t_m}.
\end{align}
Eq.\ \eqref{power} is written as an inequality because the cycle time (sum of times associated with measurement, work extraction and restoration) is larger than $t_m$. Above we argued that the measurement time can be as short as $\hbar/\Delta E$ ($10^{-12}$-$10^{-15}$s in molecular systems). Restoration (thermal relaxation) may be also short ($\sim10^{-12}$s in condensed molecular systems), so that the bound provided by Eq.\ \eqref{power} may give a useful estimate as upper bound. 

\begin{figure}[h!!!!!]
\centering
\includegraphics[width=\linewidth]{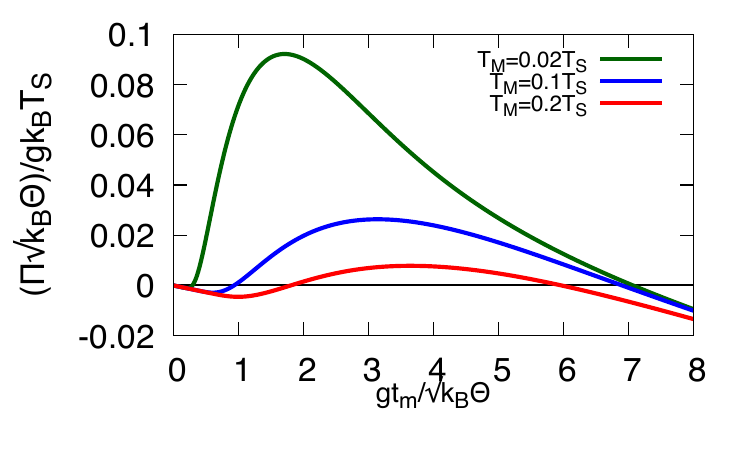}
\caption{\label{fig3} The upper bound on the output power  $\Pi(t_m)$, Eq.\ \eqref{power}, shown as a function of the system-meter interaction time, $t_m$ for different values of the meter temperature. The other parameters are the same as those used in Fig.\ \ref{fig9} (A).}
\end{figure}

Figures \ref{fig3} and \ref{figIJ} illustrate, respectively, the power output, $\Pi(t_m)$, and the efficiency, $\eta(t_m)$, as functions of the measurement time, $t_m$. Several observations can be made:

(i) To achieve a positive power output, the work output $W_{\rm out}$ must exceed the measurement cost $W_{\rm meas}$. Specifically, a lower meter temperature increases the extracted power. A reduced meter temperature narrows the initial meter distribution (as described in Eq.\ \eqref{meterIn}), enabling more information to be obtained about the system state to facilitate work extraction.

(ii) The power increases from zero, reaching a peak at intermediate measurement times before declining. This key characteristic highlights once again that information cannot be instantly acquired (even at low meter temperatures) to extract net work.

(iii) In the IE model analyzed, where the system-meter interaction remains constant until cutoff at time $t_m$, the saturation in information gain over time (see Fig.\ \ref{fig9}(B)) and the rising energy cost result in negative power output as $t_m$ approaches infinity.

(iv) Both power and efficiency indicate better IE performance at lower meter temperature, in accordance with expectations for higher quality measurement obtained using a better defined meter state, i.e., a narrower width of the initial meter distribution, Eq.\ \eqref{meterIn}.

(v) The efficiency (Fig.\ \ref{figIJ}) peaks at intermediate measurement times, necessary for the IE operation to gain information.

(vi) In Fig.\ \ref{figIJ} we also show the efficiencies associated with the total work that include in addition to the ergotropy also the Carnot thermalization work. Recall that that their a equal to $T_{ \rm S} I$, as shown in Eq.\ \eqref{integral7}. This provides an upper bound on the efficiency associated with our IE model.

As performance indicators, efficiency $\eta$ and power $\Pi$ typically offer complementary perspectives on machine operations. In conventional heat engines, maximum efficiency occurs at zero power. This behavior is also observed for the efficiency using the theoretical maximum extractable work $W_{\rm out}= W_{\rm tot}$ in Eq.\ \eqref{efficiency} whose maximal value is at $t_m=0$ (black and blue lines in Fig.\ \ref{figIJ}) where the power is zero (Fig.\ \ref{fig3}). Nonetheless, in realistic work extraction scenarios, despite smaller magnitudes, there are operation times where both efficiency and power peak simultaneously.

\begin{figure}
\centering
\includegraphics[width=1.05\linewidth]{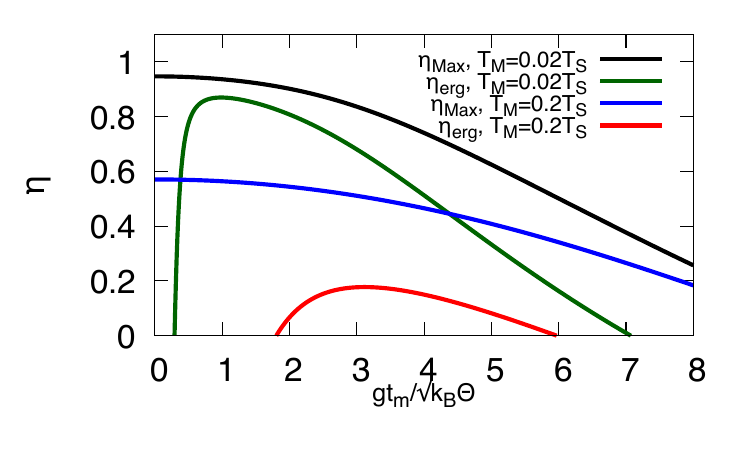}
\caption{\label{figIJ} Efficiency $\eta_{\rm erg}$ by using ergotropy, Eq.\ \eqref{ergotropy2}, for $W_{\rm out}(t_m)=Q_{\rm S}(t_m)=W_{\rm erg}(t_m)$ in Eq.\ \eqref{efficiency}, and $\eta_{\rm max}$ by using the upper bound on the extracted work, Eq.\ \eqref{integral7}, for $W_{\rm out}(t_m)=Q_{\rm S}(t_m)=T_{ \rm S} I(t_m)$ in Eq.\ \eqref{efficiency}, shown as functions of the system-meter interaction time $t_m$ for different values of the meter temperature. The parameters used are $T_{ \rm S}=300$K, $k_B\Theta=1$meV and $\Delta E=k_B T_{\rm S}$.}
\end{figure}

\section{Conclusions} 
\label{conclusion}

We have analyzed the operation of a quantum information engine that transforms heat into useful work by exploiting quantum measurement. For energy conversion devices operating at the nanoscale, the time and energy cost of the observations needed for feedback control become important aspects of the engine's performance and must be considered. Specifically, we have examined the role of measurement time (during which the system and meter are coupled) and the cost (work needed to couple and decouple the system and meter) when acquiring the information used to extract useful work. 

Because the details of the information engine's characteristics depend on the information acquisition process used, the measurement time sets a lower bound for the operation time. This time, together with the measurement energy cost are crucial for estimating standard performance metrics such as engine efficiency and power. 

In a specific example of an information engine, we have explored the role of ergotropy in providing an estimate of the measurement-enhanced extracted energy and have described a potentially practical route for extracting this work. \textit{Additionally, we found that the total extractable work associated with the measurement process, including the maximal work obtainable during the subsequent restoration of the system to its initial thermal state, is equal to the mutual information between the system and the meter multiplied by the system's temperature.} This consideration makes it possible to optimize the trade-off between engine power and efficiency.

It is interesting to note that a simple variant of our model can be used as a setup for a measurement-enhanced photochemical process. In this variant, $a$ and $b=1-a$ are the thermal probabilities for a molecule to be in the ground and a (reactive) excited state and $p'$ is chosen to ensure that the ground state population is larger than the equilibrium population $ a $. If a photon is sent only when the probability exceeds $ p' $, the photochemical yield per incident photon will be greater than the yield when photons are used indiscriminately.

Moving forward, it will be important to investigate other information engine models with different working and measurement protocols, especially in terms of studying the impact of measurement accuracy. More crucially, we must address fundamental issues concerning the finite-time operation of information-enhanced processes, such as: What is the maximum amount of information that can be extracted by observing a system during a given time period? Given a measurement-determined state or distribution of a system, what is the maximum amount of work that can be extracted from this system in a given time interval? These and similar questions will be the subject of future research.

\begin{acknowledgments}
We thank Juliette Monsel for useful comments on the manuscript. The research of H.K. is supported by the European Research Council (ERC) under the European Union’s Horizon Europe research and innovation program (101088169/NanoRecycle). The research of A.N. is supported by the Air Force Office of Scientific Research
under award number FA9550-23-1-0368 and the University of Pennsylvania.
\end{acknowledgments}

\bibliography{MS}

\begin{thebibliography}{45}%
\makeatletter
\providecommand \@ifxundefined [1]{%
 \@ifx{#1\undefined}
}%
\providecommand \@ifnum [1]{%
 \ifnum #1\expandafter \@firstoftwo
 \else \expandafter \@secondoftwo
 \fi
}%
\providecommand \@ifx [1]{%
 \ifx #1\expandafter \@firstoftwo
 \else \expandafter \@secondoftwo
 \fi
}%
\providecommand \natexlab [1]{#1}%
\providecommand \enquote  [1]{``#1''}%
\providecommand \bibnamefont  [1]{#1}%
\providecommand \bibfnamefont [1]{#1}%
\providecommand \citenamefont [1]{#1}%
\providecommand \href@noop [0]{\@secondoftwo}%
\providecommand \href [0]{\begingroup \@sanitize@url \@href}%
\providecommand \@href[1]{\@@startlink{#1}\@@href}%
\providecommand \@@href[1]{\endgroup#1\@@endlink}%
\providecommand \@sanitize@url [0]{\catcode `\\12\catcode `\$12\catcode
  `\&12\catcode `\#12\catcode `\^12\catcode `\_12\catcode `\%12\relax}%
\providecommand \@@startlink[1]{}%
\providecommand \@@endlink[0]{}%
\providecommand \url  [0]{\begingroup\@sanitize@url \@url }%
\providecommand \@url [1]{\endgroup\@href {#1}{\urlprefix }}%
\providecommand \urlprefix  [0]{URL }%
\providecommand \Eprint [0]{\href }%
\providecommand \doibase [0]{https://doi.org/}%
\providecommand \selectlanguage [0]{\@gobble}%
\providecommand \bibinfo  [0]{\@secondoftwo}%
\providecommand \bibfield  [0]{\@secondoftwo}%
\providecommand \translation [1]{[#1]}%
\providecommand \BibitemOpen [0]{}%
\providecommand \bibitemStop [0]{}%
\providecommand \bibitemNoStop [0]{.\EOS\space}%
\providecommand \EOS [0]{\spacefactor3000\relax}%
\providecommand \BibitemShut  [1]{\csname bibitem#1\endcsname}%
\let\auto@bib@innerbib\@empty
%</preamble>
\bibitem [{\citenamefont {Sagawa}\ and\ \citenamefont {Ueda}(2012)}]{sag2012}%
  \BibitemOpen
  \bibfield  {author} {\bibinfo {author} {\bibfnamefont {T.}~\bibnamefont
  {Sagawa}}\ and\ \bibinfo {author} {\bibfnamefont {M.}~\bibnamefont {Ueda}},\
  }\bibfield  {title} {\bibinfo {title} {Fluctuation theorem with information
  exchange: Role of correlations in stochastic thermodynamics},\ }\href
  {https://doi.org/10.1103/PhysRevLett.109.180602} {\bibfield  {journal}
  {\bibinfo  {journal} {Phys. Rev. Lett.}\ }\textbf {\bibinfo {volume} {109}},\
  \bibinfo {pages} {180602} (\bibinfo {year} {2012})}\BibitemShut {NoStop}%
\bibitem [{\citenamefont {Vidrighin}\ \emph {et~al.}(2016)\citenamefont
  {Vidrighin}, \citenamefont {Dahlsten}, \citenamefont {Barbieri},
  \citenamefont {Kim}, \citenamefont {Vedral},\ and\ \citenamefont
  {Walmsley}}]{vid2016}%
  \BibitemOpen
  \bibfield  {author} {\bibinfo {author} {\bibfnamefont {M.~D.}\ \bibnamefont
  {Vidrighin}}, \bibinfo {author} {\bibfnamefont {O.}~\bibnamefont {Dahlsten}},
  \bibinfo {author} {\bibfnamefont {M.}~\bibnamefont {Barbieri}}, \bibinfo
  {author} {\bibfnamefont {M.~S.}\ \bibnamefont {Kim}}, \bibinfo {author}
  {\bibfnamefont {V.}~\bibnamefont {Vedral}},\ and\ \bibinfo {author}
  {\bibfnamefont {I.~A.}\ \bibnamefont {Walmsley}},\ }\bibfield  {title}
  {\bibinfo {title} {Photonic maxwell's demon},\ }\href
  {https://doi.org/10.1103/PhysRevLett.116.050401} {\bibfield  {journal}
  {\bibinfo  {journal} {Phys. Rev. Lett.}\ }\textbf {\bibinfo {volume} {116}},\
  \bibinfo {pages} {050401} (\bibinfo {year} {2016})}\BibitemShut {NoStop}%
\bibitem [{\citenamefont {Cottet}\ \emph {et~al.}(2017)\citenamefont {Cottet},
  \citenamefont {Jezouin}, \citenamefont {Bretheau}, \citenamefont
  {Campagne-Ibarcq}, \citenamefont {Ficheux}, \citenamefont {Anders},
  \citenamefont {Auffèves}, \citenamefont {Azouit}, \citenamefont {Rouchon},\
  and\ \citenamefont {Huard}}]{cot2017}%
  \BibitemOpen
  \bibfield  {author} {\bibinfo {author} {\bibfnamefont {N.}~\bibnamefont
  {Cottet}}, \bibinfo {author} {\bibfnamefont {S.~J.}\ \bibnamefont {Jezouin}},
  \bibinfo {author} {\bibfnamefont {L.}~\bibnamefont {Bretheau}}, \bibinfo
  {author} {\bibfnamefont {P.}~\bibnamefont {Campagne-Ibarcq}}, \bibinfo
  {author} {\bibfnamefont {Q.}~\bibnamefont {Ficheux}}, \bibinfo {author}
  {\bibfnamefont {J.}~\bibnamefont {Anders}}, \bibinfo {author} {\bibfnamefont
  {A.}~\bibnamefont {Auffèves}}, \bibinfo {author} {\bibfnamefont
  {R.}~\bibnamefont {Azouit}}, \bibinfo {author} {\bibfnamefont
  {P.}~\bibnamefont {Rouchon}},\ and\ \bibinfo {author} {\bibfnamefont
  {B.}~\bibnamefont {Huard}},\ }\bibfield  {title} {\bibinfo {title} {Observing
  a quantum maxwell demon at work},\ }\href
  {https://doi.org/doi.org/10.1073/pnas.1704827114} {\bibfield  {journal}
  {\bibinfo  {journal} {Natl. Acad. Sci. U.S.A.}\ }\textbf {\bibinfo {volume}
  {114}},\ \bibinfo {pages} {7561} (\bibinfo {year} {2017})}\BibitemShut
  {NoStop}%
\bibitem [{\citenamefont {Cottet}\ and\ \citenamefont
  {Huard}(2018)}]{Cottet2018}%
  \BibitemOpen
  \bibfield  {author} {\bibinfo {author} {\bibfnamefont {N.}~\bibnamefont
  {Cottet}}\ and\ \bibinfo {author} {\bibfnamefont {B.}~\bibnamefont {Huard}},\
  }\bibinfo {title} {Maxwell's demon in superconducting circuits},\ in\ \href
  {https://doi.org/10.1007/978-3-319-99046-0_40} {\emph {\bibinfo {booktitle}
  {Thermodynamics in the Quantum Regime: Fundamental Aspects and New
  Directions}}},\ \bibinfo {editor} {edited by\ \bibinfo {editor}
  {\bibfnamefont {F.}~\bibnamefont {Binder}}, \bibinfo {editor} {\bibfnamefont
  {L.~A.}\ \bibnamefont {Correa}}, \bibinfo {editor} {\bibfnamefont
  {C.}~\bibnamefont {Gogolin}}, \bibinfo {editor} {\bibfnamefont
  {J.}~\bibnamefont {Anders}},\ and\ \bibinfo {editor} {\bibfnamefont
  {G.}~\bibnamefont {Adesso}}}\ (\bibinfo  {publisher} {Springer International
  Publishing},\ \bibinfo {address} {Cham},\ \bibinfo {year} {2018})\ pp.\
  \bibinfo {pages} {959--981}\BibitemShut {NoStop}%
\bibitem [{\citenamefont {Elouard}\ \emph {et~al.}(2017)\citenamefont
  {Elouard}, \citenamefont {Herrera-Mart\'{\i}}, \citenamefont {Huard},\ and\
  \citenamefont {Auff\`eves}}]{elo2017}%
  \BibitemOpen
  \bibfield  {author} {\bibinfo {author} {\bibfnamefont {C.}~\bibnamefont
  {Elouard}}, \bibinfo {author} {\bibfnamefont {D.}~\bibnamefont
  {Herrera-Mart\'{\i}}}, \bibinfo {author} {\bibfnamefont {B.}~\bibnamefont
  {Huard}},\ and\ \bibinfo {author} {\bibfnamefont {A.}~\bibnamefont
  {Auff\`eves}},\ }\bibfield  {title} {\bibinfo {title} {Extracting work from
  quantum measurement in maxwell's demon engines},\ }\href
  {https://doi.org/10.1103/PhysRevLett.118.260603} {\bibfield  {journal}
  {\bibinfo  {journal} {Phys. Rev. Lett.}\ }\textbf {\bibinfo {volume} {118}},\
  \bibinfo {pages} {260603} (\bibinfo {year} {2017})}\BibitemShut {NoStop}%
\bibitem [{\citenamefont {Elouard}\ and\ \citenamefont
  {Jordan}(2018)}]{elo2018}%
  \BibitemOpen
  \bibfield  {author} {\bibinfo {author} {\bibfnamefont {C.}~\bibnamefont
  {Elouard}}\ and\ \bibinfo {author} {\bibfnamefont {A.~N.}\ \bibnamefont
  {Jordan}},\ }\bibfield  {title} {\bibinfo {title} {Efficient quantum
  measurement engines},\ }\href
  {https://doi.org/10.1103/PhysRevLett.120.260601} {\bibfield  {journal}
  {\bibinfo  {journal} {Phys. Rev. Lett.}\ }\textbf {\bibinfo {volume} {120}},\
  \bibinfo {pages} {260601} (\bibinfo {year} {2018})}\BibitemShut {NoStop}%
\bibitem [{\citenamefont {Monsel}\ \emph {et~al.}(2020)\citenamefont {Monsel},
  \citenamefont {Fellous-Asiani}, \citenamefont {Huard},\ and\ \citenamefont
  {Auff\`eves}}]{mon2020}%
  \BibitemOpen
  \bibfield  {author} {\bibinfo {author} {\bibfnamefont {J.}~\bibnamefont
  {Monsel}}, \bibinfo {author} {\bibfnamefont {M.}~\bibnamefont
  {Fellous-Asiani}}, \bibinfo {author} {\bibfnamefont {B.}~\bibnamefont
  {Huard}},\ and\ \bibinfo {author} {\bibfnamefont {A.}~\bibnamefont
  {Auff\`eves}},\ }\bibfield  {title} {\bibinfo {title} {The energetic cost of
  work extraction},\ }\href {https://doi.org/10.1103/PhysRevLett.124.130601}
  {\bibfield  {journal} {\bibinfo  {journal} {Phys. Rev. Lett.}\ }\textbf
  {\bibinfo {volume} {124}},\ \bibinfo {pages} {130601} (\bibinfo {year}
  {2020})}\BibitemShut {NoStop}%
\bibitem [{\citenamefont {Bresque}\ \emph {et~al.}(2021)\citenamefont
  {Bresque}, \citenamefont {Camati}, \citenamefont {Rogers}, \citenamefont
  {Murch}, \citenamefont {Jordan},\ and\ \citenamefont {Auff\`eves}}]{bre2021}%
  \BibitemOpen
  \bibfield  {author} {\bibinfo {author} {\bibfnamefont {L.}~\bibnamefont
  {Bresque}}, \bibinfo {author} {\bibfnamefont {P.~A.}\ \bibnamefont {Camati}},
  \bibinfo {author} {\bibfnamefont {S.}~\bibnamefont {Rogers}}, \bibinfo
  {author} {\bibfnamefont {K.}~\bibnamefont {Murch}}, \bibinfo {author}
  {\bibfnamefont {A.~N.}\ \bibnamefont {Jordan}},\ and\ \bibinfo {author}
  {\bibfnamefont {A.}~\bibnamefont {Auff\`eves}},\ }\bibfield  {title}
  {\bibinfo {title} {Two-qubit engine fueled by entanglement and local
  measurements},\ }\href {https://doi.org/10.1103/PhysRevLett.126.120605}
  {\bibfield  {journal} {\bibinfo  {journal} {Phys. Rev. Lett.}\ }\textbf
  {\bibinfo {volume} {126}},\ \bibinfo {pages} {120605} (\bibinfo {year}
  {2021})}\BibitemShut {NoStop}%
\bibitem [{\citenamefont {Maxwell}(1871)}]{MaxwellBook}%
  \BibitemOpen
  \bibfield  {author} {\bibinfo {author} {\bibfnamefont {J.~C.}\ \bibnamefont
  {Maxwell}},\ }\href@noop {} {\emph {\bibinfo {title} {Theory of Heat}}}\
  (\bibinfo  {publisher} {Longmans},\ \bibinfo {address} {London},\ \bibinfo
  {year} {1871})\BibitemShut {NoStop}%
\bibitem [{\citenamefont {Zurek}(1986)}]{ZurBook}%
  \BibitemOpen
  \bibfield  {author} {\bibinfo {author} {\bibfnamefont {W.}~\bibnamefont
  {Zurek}},\ }in\ \href
  {https://doi.org/https://doi.org/10.1007/978-1-4613-2181-1} {\emph {\bibinfo
  {booktitle} {Frontiers of Nonequilibrium Statistical Physics}}},\ \bibinfo
  {series} {Nato Sci. Series B}, Vol.\ \bibinfo {volume} {135},\ \bibinfo
  {editor} {edited by\ \bibinfo {editor} {\bibfnamefont {G.}~\bibnamefont
  {Moore}}\ and\ \bibinfo {editor} {\bibfnamefont {M.~O.}\ \bibnamefont
  {Scully}}}\ (\bibinfo  {publisher} {Plenum, New York},\ \bibinfo {year}
  {1986})\ p.\ \bibinfo {pages} {145}\BibitemShut {NoStop}%
\bibitem [{\citenamefont {Maruyama}\ \emph {et~al.}(2009)\citenamefont
  {Maruyama}, \citenamefont {Nori},\ and\ \citenamefont {Vedral}}]{mar2009}%
  \BibitemOpen
  \bibfield  {author} {\bibinfo {author} {\bibfnamefont {K.}~\bibnamefont
  {Maruyama}}, \bibinfo {author} {\bibfnamefont {F.}~\bibnamefont {Nori}},\
  and\ \bibinfo {author} {\bibfnamefont {V.}~\bibnamefont {Vedral}},\
  }\bibfield  {title} {\bibinfo {title} {Colloquium: The physics of maxwell's
  demon and information},\ }\href {https://doi.org/10.1103/RevModPhys.81.1}
  {\bibfield  {journal} {\bibinfo  {journal} {Rev. Mod. Phys.}\ }\textbf
  {\bibinfo {volume} {81}},\ \bibinfo {pages} {1} (\bibinfo {year}
  {2009})}\BibitemShut {NoStop}%
\bibitem [{\citenamefont {Mandal}\ \emph {et~al.}(2013)\citenamefont {Mandal},
  \citenamefont {Quan},\ and\ \citenamefont {Jarzynski}}]{man2012}%
  \BibitemOpen
  \bibfield  {author} {\bibinfo {author} {\bibfnamefont {D.}~\bibnamefont
  {Mandal}}, \bibinfo {author} {\bibfnamefont {H.~T.}\ \bibnamefont {Quan}},\
  and\ \bibinfo {author} {\bibfnamefont {C.}~\bibnamefont {Jarzynski}},\
  }\bibfield  {title} {\bibinfo {title} {Maxwell's refrigerator: An exactly
  solvable model},\ }\href {https://doi.org/10.1103/PhysRevLett.111.030602}
  {\bibfield  {journal} {\bibinfo  {journal} {Phys. Rev. Lett.}\ }\textbf
  {\bibinfo {volume} {111}},\ \bibinfo {pages} {030602} (\bibinfo {year}
  {2013})}\BibitemShut {NoStop}%
\bibitem [{\citenamefont {Deffner}\ and\ \citenamefont
  {Jarzynski}(2013)}]{def2013}%
  \BibitemOpen
  \bibfield  {author} {\bibinfo {author} {\bibfnamefont {S.}~\bibnamefont
  {Deffner}}\ and\ \bibinfo {author} {\bibfnamefont {C.}~\bibnamefont
  {Jarzynski}},\ }\bibfield  {title} {\bibinfo {title} {Information processing
  and the second law of thermodynamics: An inclusive, hamiltonian approach},\
  }\href {https://doi.org/10.1103/PhysRevX.3.041003} {\bibfield  {journal}
  {\bibinfo  {journal} {Phys. Rev. X}\ }\textbf {\bibinfo {volume} {3}},\
  \bibinfo {pages} {041003} (\bibinfo {year} {2013})}\BibitemShut {NoStop}%
\bibitem [{\citenamefont {Strasberg}\ \emph {et~al.}(2013)\citenamefont
  {Strasberg}, \citenamefont {Schaller}, \citenamefont {Brandes},\ and\
  \citenamefont {Esposito}}]{str2013}%
  \BibitemOpen
  \bibfield  {author} {\bibinfo {author} {\bibfnamefont {P.}~\bibnamefont
  {Strasberg}}, \bibinfo {author} {\bibfnamefont {G.}~\bibnamefont {Schaller}},
  \bibinfo {author} {\bibfnamefont {T.}~\bibnamefont {Brandes}},\ and\ \bibinfo
  {author} {\bibfnamefont {M.}~\bibnamefont {Esposito}},\ }\bibfield  {title}
  {\bibinfo {title} {Thermodynamics of a physical model implementing a maxwell
  demon},\ }\href {https://doi.org/10.1103/PhysRevLett.110.040601} {\bibfield
  {journal} {\bibinfo  {journal} {Phys. Rev. Lett.}\ }\textbf {\bibinfo
  {volume} {110}},\ \bibinfo {pages} {040601} (\bibinfo {year}
  {2013})}\BibitemShut {NoStop}%
\bibitem [{\citenamefont {Horowitz}\ \emph {et~al.}(2013)\citenamefont
  {Horowitz}, \citenamefont {Sagawa},\ and\ \citenamefont
  {Parrondo}}]{hor2013}%
  \BibitemOpen
  \bibfield  {author} {\bibinfo {author} {\bibfnamefont {J.~M.}\ \bibnamefont
  {Horowitz}}, \bibinfo {author} {\bibfnamefont {T.}~\bibnamefont {Sagawa}},\
  and\ \bibinfo {author} {\bibfnamefont {J.~M.~R.}\ \bibnamefont {Parrondo}},\
  }\bibfield  {title} {\bibinfo {title} {Imitating chemical motors with optimal
  information motors},\ }\href {https://doi.org/10.1103/PhysRevLett.111.010602}
  {\bibfield  {journal} {\bibinfo  {journal} {Phys. Rev. Lett.}\ }\textbf
  {\bibinfo {volume} {111}},\ \bibinfo {pages} {010602} (\bibinfo {year}
  {2013})}\BibitemShut {NoStop}%
\bibitem [{\citenamefont {Barato}\ and\ \citenamefont
  {Seifert}(2013)}]{bar2013}%
  \BibitemOpen
  \bibfield  {author} {\bibinfo {author} {\bibfnamefont {A.~C.}\ \bibnamefont
  {Barato}}\ and\ \bibinfo {author} {\bibfnamefont {U.}~\bibnamefont
  {Seifert}},\ }\bibfield  {title} {\bibinfo {title} {An autonomous and
  reversible maxwell's demon},\ }\href
  {https://doi.org/10.1209/0295-5075/101/60001} {\bibfield  {journal} {\bibinfo
   {journal} {Europhysics Letters}\ }\textbf {\bibinfo {volume} {101}},\
  \bibinfo {pages} {60001} (\bibinfo {year} {2013})}\BibitemShut {NoStop}%
\bibitem [{\citenamefont {Landauer}(1961)}]{lan1961}%
  \BibitemOpen
  \bibfield  {author} {\bibinfo {author} {\bibfnamefont {R.}~\bibnamefont
  {Landauer}},\ }\bibfield  {title} {\bibinfo {title} {Irreversibility and heat
  generation in the computing process},\ }\href
  {https://doi.org/10.1147/rd.53.0183} {\bibfield  {journal} {\bibinfo
  {journal} {IBM Journal of Research and Development}\ }\textbf {\bibinfo
  {volume} {5}},\ \bibinfo {pages} {183} (\bibinfo {year} {1961})}\BibitemShut
  {NoStop}%
\bibitem [{\citenamefont {de~Oliveira~Junior}\ \emph
  {et~al.}(2024)\citenamefont {de~Oliveira~Junior}, \citenamefont {Brask},\
  and\ \citenamefont {Lipka-Bartosik}}]{jun2024}%
  \BibitemOpen
  \bibfield  {author} {\bibinfo {author} {\bibfnamefont {A.}~\bibnamefont
  {de~Oliveira~Junior}}, \bibinfo {author} {\bibfnamefont {J.~B.}\ \bibnamefont
  {Brask}},\ and\ \bibinfo {author} {\bibfnamefont {P.}~\bibnamefont
  {Lipka-Bartosik}},\ }\bibfield  {title} {\bibinfo {title} {Heat as a witness
  of quantum properties},\ }\href {https://arxiv.org/abs/2408.06418} {\
  (\bibinfo {year} {2024})},\ \Eprint {https://arxiv.org/abs/2408.06418}
  {arXiv:2408.06418 [quant-ph]} \BibitemShut {NoStop}%
\bibitem [{\citenamefont {Von~Neumann}(1955)}]{VonNeumannBook}%
  \BibitemOpen
  \bibfield  {author} {\bibinfo {author} {\bibfnamefont {J.}~\bibnamefont
  {Von~Neumann}},\ }\href@noop {} {\emph {\bibinfo {title} {Mathematical
  Foundations of Quantum Mechanics}}}\ (\bibinfo  {publisher} {Princeton Univ.
  Press.},\ \bibinfo {address} {Princeton, NJ},\ \bibinfo {year}
  {1955})\BibitemShut {NoStop}%
\bibitem [{\citenamefont {Guryanova}\ \emph {et~al.}(2020)\citenamefont
  {Guryanova}, \citenamefont {Friis},\ and\ \citenamefont {Huber}}]{gur2020}%
  \BibitemOpen
  \bibfield  {author} {\bibinfo {author} {\bibfnamefont {Y.}~\bibnamefont
  {Guryanova}}, \bibinfo {author} {\bibfnamefont {N.}~\bibnamefont {Friis}},\
  and\ \bibinfo {author} {\bibfnamefont {M.}~\bibnamefont {Huber}},\ }\bibfield
   {title} {\bibinfo {title} {Ideal {P}rojective {M}easurements {H}ave
  {I}nfinite {R}esource {C}osts},\ }\href
  {https://doi.org/10.22331/q-2020-01-13-222} {\bibfield  {journal} {\bibinfo
  {journal} {{Quantum}}\ }\textbf {\bibinfo {volume} {4}},\ \bibinfo {pages}
  {222} (\bibinfo {year} {2020})}\BibitemShut {NoStop}%
\bibitem [{\citenamefont {Taranto}\ \emph {et~al.}(2023)\citenamefont
  {Taranto}, \citenamefont {Bakhshinezhad}, \citenamefont {Bluhm},
  \citenamefont {Silva}, \citenamefont {Friis}, \citenamefont {Lock},
  \citenamefont {Vitagliano}, \citenamefont {Binder}, \citenamefont {Debarba},
  \citenamefont {Schwarzhans}, \citenamefont {Clivaz},\ and\ \citenamefont
  {Huber}}]{tar2023}%
  \BibitemOpen
  \bibfield  {author} {\bibinfo {author} {\bibfnamefont {P.}~\bibnamefont
  {Taranto}}, \bibinfo {author} {\bibfnamefont {F.}~\bibnamefont
  {Bakhshinezhad}}, \bibinfo {author} {\bibfnamefont {A.}~\bibnamefont
  {Bluhm}}, \bibinfo {author} {\bibfnamefont {R.}~\bibnamefont {Silva}},
  \bibinfo {author} {\bibfnamefont {N.}~\bibnamefont {Friis}}, \bibinfo
  {author} {\bibfnamefont {M.~P.}\ \bibnamefont {Lock}}, \bibinfo {author}
  {\bibfnamefont {G.}~\bibnamefont {Vitagliano}}, \bibinfo {author}
  {\bibfnamefont {F.~C.}\ \bibnamefont {Binder}}, \bibinfo {author}
  {\bibfnamefont {T.}~\bibnamefont {Debarba}}, \bibinfo {author} {\bibfnamefont
  {E.}~\bibnamefont {Schwarzhans}}, \bibinfo {author} {\bibfnamefont
  {F.}~\bibnamefont {Clivaz}},\ and\ \bibinfo {author} {\bibfnamefont
  {M.}~\bibnamefont {Huber}},\ }\bibfield  {title} {\bibinfo {title} {Landauer
  versus nernst: What is the true cost of cooling a quantum system?},\ }\href
  {https://doi.org/10.1103/PRXQuantum.4.010332} {\bibfield  {journal} {\bibinfo
   {journal} {PRX Quantum}\ }\textbf {\bibinfo {volume} {4}},\ \bibinfo {pages}
  {010332} (\bibinfo {year} {2023})}\BibitemShut {NoStop}%
\bibitem [{\citenamefont {Mensky}(2000)}]{MenskyBook}%
  \BibitemOpen
  \bibfield  {author} {\bibinfo {author} {\bibfnamefont {M.~M.}\ \bibnamefont
  {Mensky}},\ }\href@noop {} {\emph {\bibinfo {title} {Quantum Mesurements and
  Decoherence}}}\ (\bibinfo  {publisher} {Kluwer Academic Publishers},\
  \bibinfo {address} {Dordrecht, The Netherlands},\ \bibinfo {year}
  {2000})\BibitemShut {NoStop}%
\bibitem [{\citenamefont {Jacobs}\ and\ \citenamefont {Steck}(2006)}]{jac2006}%
  \BibitemOpen
  \bibfield  {author} {\bibinfo {author} {\bibfnamefont {K.}~\bibnamefont
  {Jacobs}}\ and\ \bibinfo {author} {\bibfnamefont {D.~A.}\ \bibnamefont
  {Steck}},\ }\bibfield  {title} {\bibinfo {title} {A straightforward
  introduction to continuous quantum measurement},\ }\href
  {https://doi.org/10.1080/00107510601101934} {\bibfield  {journal} {\bibinfo
  {journal} {Contemporary Physics}\ }\textbf {\bibinfo {volume} {47}},\
  \bibinfo {pages} {279} (\bibinfo {year} {2006})}\BibitemShut {NoStop}%
\bibitem [{\citenamefont {Annby-Andersson}\ \emph {et~al.}(2022)\citenamefont
  {Annby-Andersson}, \citenamefont {Bakhshinezhad}, \citenamefont
  {Bhattacharyya}, \citenamefont {De~Sousa}, \citenamefont {Jarzynski},
  \citenamefont {Samuelsson},\ and\ \citenamefont {Potts}}]{ann2022}%
  \BibitemOpen
  \bibfield  {author} {\bibinfo {author} {\bibfnamefont {B.}~\bibnamefont
  {Annby-Andersson}}, \bibinfo {author} {\bibfnamefont {F.}~\bibnamefont
  {Bakhshinezhad}}, \bibinfo {author} {\bibfnamefont {D.}~\bibnamefont
  {Bhattacharyya}}, \bibinfo {author} {\bibfnamefont {G.}~\bibnamefont
  {De~Sousa}}, \bibinfo {author} {\bibfnamefont {C.}~\bibnamefont {Jarzynski}},
  \bibinfo {author} {\bibfnamefont {P.}~\bibnamefont {Samuelsson}},\ and\
  \bibinfo {author} {\bibfnamefont {P.~P.}\ \bibnamefont {Potts}},\ }\bibfield
  {title} {\bibinfo {title} {Quantum fokker-planck master equation for
  continuous feedback control},\ }\href
  {https://doi.org/10.1103/PhysRevLett.129.050401} {\bibfield  {journal}
  {\bibinfo  {journal} {Phys. Rev. Lett.}\ }\textbf {\bibinfo {volume} {129}},\
  \bibinfo {pages} {050401} (\bibinfo {year} {2022})}\BibitemShut {NoStop}%
\bibitem [{\citenamefont {Ochoa}\ \emph {et~al.}(2018)\citenamefont {Ochoa},
  \citenamefont {Belzig},\ and\ \citenamefont {Nitzan}}]{och2018}%
  \BibitemOpen
  \bibfield  {author} {\bibinfo {author} {\bibfnamefont {M.}~\bibnamefont
  {Ochoa}}, \bibinfo {author} {\bibfnamefont {W.}~\bibnamefont {Belzig}},\ and\
  \bibinfo {author} {\bibfnamefont {A.}~\bibnamefont {Nitzan}},\ }\bibfield
  {title} {\bibinfo {title} {Simultaneous weak measurement of non-commuting
  observables: a generalized arthurs-kelly protocol},\ }\href
  {https://doi.org/10.1038/s41598-018-33562-0} {\bibfield  {journal} {\bibinfo
  {journal} {Sci Rep}\ }\textbf {\bibinfo {volume} {8}},\ \bibinfo {pages}
  {15781} (\bibinfo {year} {2018})}\BibitemShut {NoStop}%
\bibitem [{\citenamefont {Wiseman}(1994)}]{wis1994}%
  \BibitemOpen
  \bibfield  {author} {\bibinfo {author} {\bibfnamefont {H.~M.}\ \bibnamefont
  {Wiseman}},\ }\bibfield  {title} {\bibinfo {title} {Quantum theory of
  continuous feedback},\ }\href {https://doi.org/10.1103/PhysRevA.49.2133}
  {\bibfield  {journal} {\bibinfo  {journal} {Phys. Rev. A}\ }\textbf {\bibinfo
  {volume} {49}},\ \bibinfo {pages} {2133} (\bibinfo {year}
  {1994})}\BibitemShut {NoStop}%
\bibitem [{\citenamefont {Sagawa}\ and\ \citenamefont {Ueda}(2010)}]{sag2010}%
  \BibitemOpen
  \bibfield  {author} {\bibinfo {author} {\bibfnamefont {T.}~\bibnamefont
  {Sagawa}}\ and\ \bibinfo {author} {\bibfnamefont {M.}~\bibnamefont {Ueda}},\
  }\bibfield  {title} {\bibinfo {title} {Generalized jarzynski equality under
  nonequilibrium feedback control},\ }\href
  {https://doi.org/10.1103/PhysRevLett.104.090602} {\bibfield  {journal}
  {\bibinfo  {journal} {Phys. Rev. Lett.}\ }\textbf {\bibinfo {volume} {104}},\
  \bibinfo {pages} {090602} (\bibinfo {year} {2010})}\BibitemShut {NoStop}%
\bibitem [{\citenamefont {Toyabe}\ \emph {et~al.}(2010)\citenamefont {Toyabe},
  \citenamefont {Sagawa}, \citenamefont {Ueda}, \citenamefont {Muneyuki},\ and\
  \citenamefont {Sano}}]{toy2010}%
  \BibitemOpen
  \bibfield  {author} {\bibinfo {author} {\bibfnamefont {S.}~\bibnamefont
  {Toyabe}}, \bibinfo {author} {\bibfnamefont {T.}~\bibnamefont {Sagawa}},
  \bibinfo {author} {\bibfnamefont {M.}~\bibnamefont {Ueda}}, \bibinfo {author}
  {\bibfnamefont {E.}~\bibnamefont {Muneyuki}},\ and\ \bibinfo {author}
  {\bibfnamefont {M.}~\bibnamefont {Sano}},\ }\bibfield  {title} {\bibinfo
  {title} {Experimental demonstration of information-to-energy conversion and
  validation of the generalized jarzynski equality},\ }\href
  {https://doi.org/10.1038/nphys1821} {\bibfield  {journal} {\bibinfo
  {journal} {Nature Phys}\ ,\ \bibinfo {pages} {988–992}} (\bibinfo {year}
  {2010})}\BibitemShut {NoStop}%
\bibitem [{\citenamefont {Ghirardi}\ \emph {et~al.}(1979)\citenamefont
  {Ghirardi}, \citenamefont {Omero}, \citenamefont {Weber},\ and\ \citenamefont
  {Rimini}}]{ghi1979}%
  \BibitemOpen
  \bibfield  {author} {\bibinfo {author} {\bibfnamefont {G.~C.}\ \bibnamefont
  {Ghirardi}}, \bibinfo {author} {\bibfnamefont {C.}~\bibnamefont {Omero}},
  \bibinfo {author} {\bibfnamefont {T.}~\bibnamefont {Weber}},\ and\ \bibinfo
  {author} {\bibfnamefont {A.}~\bibnamefont {Rimini}},\ }\bibfield  {title}
  {\bibinfo {title} {Small-time behaviour of quantum nondecay probability and
  zeno's paradox in quantum mechanics},\ }\href@noop {} {\bibfield  {journal}
  {\bibinfo  {journal} {Il Nuevo Cimento}\ }\textbf {\bibinfo {volume} {52}},\
  \bibinfo {pages} {421–442} (\bibinfo {year} {1979})}\BibitemShut {NoStop}%
\bibitem [{\citenamefont {Brand\~ao}\ \emph {et~al.}(2013)\citenamefont
  {Brand\~ao}, \citenamefont {Horodecki}, \citenamefont {Oppenheim},
  \citenamefont {Renes},\ and\ \citenamefont {Spekkens}}]{bra2013}%
  \BibitemOpen
  \bibfield  {author} {\bibinfo {author} {\bibfnamefont {F.~G. S.~L.}\
  \bibnamefont {Brand\~ao}}, \bibinfo {author} {\bibfnamefont {M.}~\bibnamefont
  {Horodecki}}, \bibinfo {author} {\bibfnamefont {J.}~\bibnamefont
  {Oppenheim}}, \bibinfo {author} {\bibfnamefont {J.~M.}\ \bibnamefont
  {Renes}},\ and\ \bibinfo {author} {\bibfnamefont {R.~W.}\ \bibnamefont
  {Spekkens}},\ }\bibfield  {title} {\bibinfo {title} {Resource theory of
  quantum states out of thermal equilibrium},\ }\href
  {https://doi.org/10.1103/PhysRevLett.111.250404} {\bibfield  {journal}
  {\bibinfo  {journal} {Phys. Rev. Lett.}\ }\textbf {\bibinfo {volume} {111}},\
  \bibinfo {pages} {250404} (\bibinfo {year} {2013})}\BibitemShut {NoStop}%
\bibitem [{\citenamefont {Bu\ss{}hardt}\ and\ \citenamefont
  {Freyberger}(2010)}]{bus2010}%
  \BibitemOpen
  \bibfield  {author} {\bibinfo {author} {\bibfnamefont {M.}~\bibnamefont
  {Bu\ss{}hardt}}\ and\ \bibinfo {author} {\bibfnamefont {M.}~\bibnamefont
  {Freyberger}},\ }\bibfield  {title} {\bibinfo {title} {Timing in quantum
  measurements of position and momentum},\ }\href
  {https://doi.org/10.1103/PhysRevA.82.042117} {\bibfield  {journal} {\bibinfo
  {journal} {Phys. Rev. A}\ }\textbf {\bibinfo {volume} {82}},\ \bibinfo
  {pages} {042117} (\bibinfo {year} {2010})}\BibitemShut {NoStop}%
\bibitem [{\citenamefont {Korbicz}\ \emph {et~al.}(2017)\citenamefont
  {Korbicz}, \citenamefont {Aguilar}, \citenamefont {\ifmmode \acute{C}\else
  \'{C}\fi{}wikli\ifmmode~\acute{n}\else \'{n}\fi{}ski},\ and\ \citenamefont
  {Horodecki}}]{kor2017}%
  \BibitemOpen
  \bibfield  {author} {\bibinfo {author} {\bibfnamefont {J.~K.}\ \bibnamefont
  {Korbicz}}, \bibinfo {author} {\bibfnamefont {E.~A.}\ \bibnamefont
  {Aguilar}}, \bibinfo {author} {\bibfnamefont {P.}~\bibnamefont {\ifmmode
  \acute{C}\else \'{C}\fi{}wikli\ifmmode~\acute{n}\else \'{n}\fi{}ski}},\ and\
  \bibinfo {author} {\bibfnamefont {P.}~\bibnamefont {Horodecki}},\ }\bibfield
  {title} {\bibinfo {title} {Generic appearance of objective results in quantum
  measurements},\ }\href {https://doi.org/10.1103/PhysRevA.96.032124}
  {\bibfield  {journal} {\bibinfo  {journal} {Phys. Rev. A}\ }\textbf {\bibinfo
  {volume} {96}},\ \bibinfo {pages} {032124} (\bibinfo {year}
  {2017})}\BibitemShut {NoStop}%
\bibitem [{\citenamefont {Hance}\ and\ \citenamefont
  {Hossenfelder}(2022)}]{han2022}%
  \BibitemOpen
  \bibfield  {author} {\bibinfo {author} {\bibfnamefont {J.~R.}\ \bibnamefont
  {Hance}}\ and\ \bibinfo {author} {\bibfnamefont {S.}~\bibnamefont
  {Hossenfelder}},\ }\bibfield  {title} {\bibinfo {title} {What does it take to
  solve the measurement problem?},\ }\href
  {https://doi.org/10.1088/2399-6528/ac96cf} {\bibfield  {journal} {\bibinfo
  {journal} {Journal of Physics Communications}\ }\textbf {\bibinfo {volume}
  {6}},\ \bibinfo {pages} {102001} (\bibinfo {year} {2022})}\BibitemShut
  {NoStop}%
\bibitem [{foo({\natexlab{a}})}]{footnote18}%
  \BibitemOpen
  \href@noop {} {\emph {\bibinfo {title} {Starting from initial thermal states
  is not a fundamental requirement. What is absolutely essential is that the
  working cycle ends with restoring the initial system and meter states, and
  thermal equilibrium represents easily restorable initial
  states.}}}\BibitemShut {Stop}%
\bibitem [{SM1()}]{SM1}%
  \BibitemOpen
  \href@noop {} {\emph {\bibinfo {title} {Supplemental Material}}}\BibitemShut
  {NoStop}%
\bibitem [{\citenamefont {Jordan}\ \emph {et~al.}(2020)\citenamefont {Jordan},
  \citenamefont {Elouard},\ and\ \citenamefont {Auffèves}}]{jor2020}%
  \BibitemOpen
  \bibfield  {author} {\bibinfo {author} {\bibfnamefont {A.}~\bibnamefont
  {Jordan}}, \bibinfo {author} {\bibfnamefont {C.}~\bibnamefont {Elouard}},\
  and\ \bibinfo {author} {\bibfnamefont {A.}~\bibnamefont {Auffèves}},\
  }\bibfield  {title} {\bibinfo {title} {Quantum measurement engines and their
  relevance for quantum interpretations},\ }\href
  {https://doi.org/doi.org/10.1007/s40509-019-00217-2} {\bibfield  {journal}
  {\bibinfo  {journal} {Quantum Stud.: Math. Found.}\ ,\ \bibinfo {pages}
  {203–215}} (\bibinfo {year} {2020})}\BibitemShut {NoStop}%
\bibitem [{\citenamefont {Atmanspacher}(1997)}]{atm1997}%
  \BibitemOpen
  \bibfield  {author} {\bibinfo {author} {\bibfnamefont {H.}~\bibnamefont
  {Atmanspacher}},\ }\bibfield  {title} {\bibinfo {title} {Cartesian cut,
  heisenberg cut, and the concept of complexity},\ }\href
  {https://doi.org/10.1080/02604027.1997.9972639} {\bibfield  {journal}
  {\bibinfo  {journal} {World Futures}\ }\textbf {\bibinfo {volume} {49}},\
  \bibinfo {pages} {333} (\bibinfo {year} {1997})}\BibitemShut {NoStop}%
\bibitem [{\citenamefont {L~Latune}\ and\ \citenamefont
  {Elouard}(2025)}]{lat2024}%
  \BibitemOpen
  \bibfield  {author} {\bibinfo {author} {\bibfnamefont {C.}~\bibnamefont
  {L~Latune}}\ and\ \bibinfo {author} {\bibfnamefont {C.}~\bibnamefont
  {Elouard}},\ }\bibfield  {title} {\bibinfo {title} {A thermodynamically
  consistent approach to the energy costs of quantum measurements},\ }\href
  {https://doi.org/10.22331/q-2025-01-28-1614} {\bibfield  {journal} {\bibinfo
  {journal} {{Quantum}}\ }\textbf {\bibinfo {volume} {9}},\ \bibinfo {pages}
  {1614} (\bibinfo {year} {2025})}\BibitemShut {NoStop}%
\bibitem [{\citenamefont {Arthurs}\ and\ \citenamefont
  {Kelly}(1965)}]{art1965}%
  \BibitemOpen
  \bibfield  {author} {\bibinfo {author} {\bibfnamefont {E.}~\bibnamefont
  {Arthurs}}\ and\ \bibinfo {author} {\bibfnamefont {J.~L.}\ \bibnamefont
  {Kelly}},\ }\bibfield  {title} {\bibinfo {title} {B.s.t.j. briefs: On the
  simultaneous measurement of a pair of conjugate observables},\ }\href
  {https://doi.org/10.1002/j.1538-7305.1965.tb01684.x} {\bibfield  {journal}
  {\bibinfo  {journal} {The Bell System Technical Journal}\ }\textbf {\bibinfo
  {volume} {44}},\ \bibinfo {pages} {725} (\bibinfo {year} {1965})}\BibitemShut
  {NoStop}%
\bibitem [{\citenamefont {Machluf}\ \emph {et~al.}(2013)\citenamefont
  {Machluf}, \citenamefont {Japha},\ and\ \citenamefont {Folman}}]{mac2013}%
  \BibitemOpen
  \bibfield  {author} {\bibinfo {author} {\bibfnamefont {S.}~\bibnamefont
  {Machluf}}, \bibinfo {author} {\bibfnamefont {Y.}~\bibnamefont {Japha}},\
  and\ \bibinfo {author} {\bibfnamefont {R.}~\bibnamefont {Folman}},\
  }\bibfield  {title} {\bibinfo {title} {Coherent stern–gerlach momentum
  splitting on an atom chip},\ }\href {https://doi.org/10.1038/ncomms3424}
  {\bibfield  {journal} {\bibinfo  {journal} {Nat Commun}\ }\textbf {\bibinfo
  {volume} {4}},\ \bibinfo {pages} {2424} (\bibinfo {year} {2013})}\BibitemShut
  {NoStop}%
\bibitem [{foo({\natexlab{b}})}]{footnote4}%
  \BibitemOpen
  \href@noop {} {\emph {\bibinfo {title} {Note that "ensemble average" here
  refers to average over the meter outcome, in contrast to the standard meaning
  as average over initial conditions.}}}\BibitemShut {Stop}%
\bibitem [{\citenamefont {Shannon}(1948)}]{sha1948}%
  \BibitemOpen
  \bibfield  {author} {\bibinfo {author} {\bibfnamefont {C.~E.}\ \bibnamefont
  {Shannon}},\ }\bibfield  {title} {\bibinfo {title} {A mathematical theory of
  communication},\ }\href {https://doi.org/10.1002/j.1538-7305.1948.tb01338.x}
  {\bibfield  {journal} {\bibinfo  {journal} {The Bell System Technical
  Journal}\ }\textbf {\bibinfo {volume} {27}},\ \bibinfo {pages} {379}
  (\bibinfo {year} {1948})}\BibitemShut {NoStop}%
\bibitem [{\citenamefont {Allahverdyan}\ \emph {et~al.}(2004)\citenamefont
  {Allahverdyan}, \citenamefont {Balian},\ and\ \citenamefont
  {Nieuwenhuizen}}]{all2004}%
  \BibitemOpen
  \bibfield  {author} {\bibinfo {author} {\bibfnamefont {A.~E.}\ \bibnamefont
  {Allahverdyan}}, \bibinfo {author} {\bibfnamefont {R.}~\bibnamefont
  {Balian}},\ and\ \bibinfo {author} {\bibfnamefont {T.~M.}\ \bibnamefont
  {Nieuwenhuizen}},\ }\bibfield  {title} {\bibinfo {title} {Maximal work
  extraction from finite quantum systems},\ }\href
  {https://doi.org/10.1209/epl/i2004-10101-2} {\bibfield  {journal} {\bibinfo
  {journal} {Europhysics Letters}\ }\textbf {\bibinfo {volume} {67}},\ \bibinfo
  {pages} {565} (\bibinfo {year} {2004})}\BibitemShut {NoStop}%
\bibitem [{SM2()}]{SM2}%
  \BibitemOpen
  \href@noop {} {\emph {\bibinfo {title} {See Eqs.\ (S.5) and (S.6) in
  \cite{SM1}.}}}\BibitemShut {Stop}%
\bibitem [{\citenamefont {Barato}\ and\ \citenamefont
  {Seifert}(2014)}]{bar2014}%
  \BibitemOpen
  \bibfield  {author} {\bibinfo {author} {\bibfnamefont {A.~C.}\ \bibnamefont
  {Barato}}\ and\ \bibinfo {author} {\bibfnamefont {U.}~\bibnamefont
  {Seifert}},\ }\bibfield  {title} {\bibinfo {title} {Unifying three
  perspectives on information processing in stochastic thermodynamics},\ }\href
  {https://doi.org/10.1103/PhysRevLett.112.090601} {\bibfield  {journal}
  {\bibinfo  {journal} {Phys. Rev. Lett.}\ }\textbf {\bibinfo {volume} {112}},\
  \bibinfo {pages} {090601} (\bibinfo {year} {2014})}\BibitemShut {NoStop}%
\end{thebibliography}%

%%%%%%%%%%%%%%%%%%%%%%%%%%%%%%%%%%%%%%%%%%

\newpage
\appendix
\onecolumngrid
\section*{Appendix}
 In Section \ref{dynamics}, the dynamics of the coupled system-meter are investigated before calculating the expectation value of the meter state in Section \ref{momentumshort}. The mutual information between system and meter is analyzed in Section \ref{mutualinformation}. Section \ref{energyshort} examines the energy change of the meter and the measurement energy required for the measurement protocol and the possible energy extraction after the measurement. Section \ref{retherm} presents the detailed derivation of work extraction during rethermalization. Finally, Section \ref{bound} demonstrates that the total work extraction during the cycle can be written by the ergotropy plus work extraction during rethermalization.
 
\section{Dynamics of system and meter}
\label{dynamics}
We use an iterative numerical scheme to determine unitary time evolution of the density matrix for the coupled 2SS and meter given under the total Hamiltonian $\hat{H}$ given the initial density matrix (Eq.\ (2) in the main text where we define $\rho_M\equiv\ket{D}\bra{D}$ with $\braket{p|D}=D(p)=\frac{1}{(2\pi k_B T_M)^{1/4}}\exp{\big[p^2/4k_BT_M\big]}$) by
\begin{align}
\label{Dy}
 \hat{\rho}(t) &=e^{-i \hat{H}t/\hbar} \hat{\rho}(0) e^{i \hat{H}t/\hbar} \\ \notag
 &= a  \big( e^{-i \hat{H}_0 \Delta t/\hbar}e^{-i \frac{\hat{p}^2}{2} \Delta t/\hbar}e^{-i\hat{V} \Delta t /\hbar}\big)^N \ket{0}\bra{0} \otimes  \ket{D} \bra{D}   \big(e^{i \hat{H}_0 \Delta t/\hbar}e^{i \frac{\hat{p}^2}{2} \Delta t/\hbar}e^{i\hat{V} \Delta t /\hbar}\big)^N  
\\ \notag &+b \big(e^{-i\hat{H}_0 \Delta t/\hbar}e^{-i\frac{\hat{p}^2}{2} \Delta t/\hbar}e^{-i\hat{V} \Delta t /\hbar}\big)^N\ket{1} \bra{1} \otimes   \ket{D} \bra{D} \big( e^{i\hat{H}_0 \Delta t/\hbar}e^{i\frac{\hat{p}^2}{2} \Delta t/\hbar}e^{i \hat{V} \Delta t /\hbar} \big)^N ,
\end{align}
while using the Trotter-splitting $e^{i(\hat{H}_0+\frac{\hat{p}^2}{2}+\hat{V})t/\hbar}=\big( e^{i\hat{H}_0\Delta t/\hbar}e^{i\frac{\hat{p}^2}{2}\Delta t/\hbar}e^{i\hat{V}\Delta t/\hbar} \big)^N$ with $\Delta t =t/N$ for $N\to \infty$.

The joint probability $P_i(p,t)$ can be solved analytically using the Trotter splitting in Eq.\ \eqref{Dy} which results to
\begin{align}
\label{Dy}
P_i(p,t) &=\sum_i\bra{i}\bra{p} \hat{\rho}(t) \ket{p} \ket{i} =\sum_i \bra{i}\bra{p}e^{-i \hat{H}t/\hbar} \hat{\rho}(0) e^{i \hat{H}t/\hbar} \ket{p} \ket{i}  \\ \notag
 &= a \bra{D}  \bra{0} \big(e^{i \hat{H}_0 \Delta t/\hbar}e^{i \frac{\hat{p}^2}{2} \Delta t/\hbar}e^{i\hat{V} \Delta t /\hbar}\big)^N \ket{0}\bra{0} \\ \notag &\otimes\ket{p}\bra{p}  \big( e^{-i \hat{H}_0 \Delta t/\hbar} e^{-i \frac{\hat{p}^2}{2} \Delta t/\hbar}e^{-i\hat{V} \Delta t /\hbar}\big)^N \ket{0}  \ket{D}  \\ \notag &+b \bra{D}  \bra{1} \big( e^{i \hat{H}_0 \Delta t/\hbar}e^{i \frac{\hat{p}^2}{2} \Delta t/\hbar}e^{i\hat{V} \Delta t /\hbar} \big)^N \ket{1}\bra{1}  \\ \notag &\otimes \ket{p}\bra{p}  \big(e^{-i \hat{H}_0 \Delta t/\hbar} e^{-i \frac{\hat{p}^2}{2} \Delta t/\hbar}e^{-i\hat{V} \Delta t /\hbar}\big)^N\ket{1}  \ket{D} \\  \notag
    &= a |\braket{p|D}|^2 
  \\ \label{eigens} &+b \int ds \int dm\int ds' \int dm' \int dx \braket{D(s)|s} e^{i \frac{s^2}{2}\Delta t/\hbar}\braket{s|x}e^{i g x \Delta t /\hbar}\braket{x|m}
  \\ \notag
  & \bra{m}  \big(e^{i \frac{\hat{p}^2}{2} \Delta t/\hbar}e^{i g \hat{x} \Delta t /\hbar} \big)^{N-1} \ket{p}\bra{p} \big( e^{-i \frac{\hat{p}^2}{2} \Delta t/\hbar}e^{-i g \hat{x} \Delta t /\hbar}\big)^{N-1} \ket{m'} \\ \notag &\bra{m'}  e^{-i \frac{\hat{p}^2}{2} \Delta t/\hbar}\ket{s'}\bra{s'}e^{-i g \hat{x} \Delta t /\hbar} \ket{D} \\  \notag
  &= a |\braket{p|D}|^2 
  \\ \label{eigens2} &+b \braket{D|p+gN\Delta t} \Pi_{k=1}^N \bigg[ e^{i(p+g k \Delta t)^2\Delta t/2\hbar} \bigg]  \\ \notag &\Pi_{k=1}^N \bigg[ e^{-i(p+g k \Delta t)^2 \Delta t/2\hbar} \bigg] \braket{p+gN\Delta t|D} 
  \\  \notag
  &= a |D(p)|^2 +b |D(p+gN\Delta t)|^2 =a |D(p)|^2 +b |D(p+gt)|^2,
\end{align}
where we have exploit the completeness relation for the momentum eigenstates $\int dm \ket{m}\bra{m}=\int dm' \ket{m'}\bra{m'}=\int ds \ket{s}\bra{s}=\int ds' \ket{s'}\bra{s'}=\mathbb{I}$ and for the position eigenstates $\int dx \ket{x}\bra{x}=\mathbb{I}$ in line \eqref{eigens}. Using the relation $\braket{x|m}=e^{imxt/\hbar}/\sqrt{2\pi}$ and the identity $\int dx e^{i(q-a)x/\hbar}=2\pi \delta(q-a)$, one arrives iteratively to the expression in line \eqref{eigens2}.
\section{Average meter state}
\label{momentumshort}
The expectation value of the meter outcome $\langle p(t_m) \rangle$  after the measurement of duration $t_m$, while using the Trotter-splitting $e^{i(\hat{H}_0+\frac{\hat{p}^2}{2}+\hat{V})t_m/\hbar}=\big( e^{i\hat{H}_0\Delta t/\hbar}e^{i\frac{\hat{p}^2}{2}\Delta t/\hbar}e^{i\hat{V}\Delta t/\hbar} \big)^N$ with $\Delta t =t_m/N$ for $N\to \infty$, reads
\begin{align}
\label{P1}
\langle p(t_m) \rangle &= \mathrm{tr}[\hat{\rho}(t_m) \hat{p}] \\ \notag &= a \bra{D}  \bra{0} \big(e^{i \hat{H}_0 \Delta t/\hbar}e^{i \frac{\hat{p}^2}{2} \Delta t/\hbar}e^{i\hat{V} \Delta t /\hbar}\big)^N  \hat{p} \big( e^{-i \hat{H}_0 \Delta t/\hbar} e^{-i \frac{\hat{p}^2}{2} \Delta t/\hbar}e^{-i\hat{V} \Delta t /\hbar}\big)^N \ket{0}  \ket{D}  \\ \notag &+b \bra{D}  \bra{1} \big( e^{i \hat{H}_0 \Delta t/\hbar}e^{i \frac{\hat{p}^2}{2} \Delta t/\hbar}e^{i\hat{V} \Delta t /\hbar} \big)^N  \hat{p} \big(e^{-i \hat{H}_0 \Delta t/\hbar} e^{-i \frac{\hat{p}^2}{2} \Delta t/\hbar}e^{-i\hat{V} \Delta t /\hbar}\big)^N\ket{1}  \ket{D} \\  \notag
&= b g \int_{-\infty}^{\infty} dp D(p+gN\Delta t) \Pi_{k=1}^N \bigg[ e^{i(p+g k \Delta t)^2\Delta t/2\hbar} \bigg] p \Pi_{k=1}^N \bigg[ e^{-i(p+g k \Delta t)^2\Delta t/2\hbar} \bigg] D(p+gN\Delta t)\\ \notag
&=  bg  \bigg ( \frac{1}{2 \pi k_B T_M} \bigg )^{1/2}  \int_{-\infty}^{\infty} dp e^{- (p+g t_m)^2/2k_B T_M} p =-b g t_m. 
\end{align}

\section{Mutual Information}
\label{mutualinformation}

The information gain, $I(t_m)$, Eq.\ 10, in the main text, in this measurement process can be quantified by averaging the conditional system entropy
$S(t_m|p)=-k_B \sum_{i=0}^{1}  P_i(t_m|p) \ln{P_i(t_m|p)}$ over an ensemble of identical measurements by
\begin{align}
S(t_m) &= \int dp Q(p,t_m) S(t_m|p) \\ \notag &= -k_B \int_{-\infty}^\infty dp \sum_{i=0}^{1}  P_i(p,t_m) \ln{P_i(t_m|p)}
\end{align}
leading to
\begin{align}
\label{know}
I(t_m)&\equiv S(0)-S(t_m) \\
&=-k_B \int_{-\infty}^\infty dp \sum_{i=0}^{1}  P_i(p,0) \ln{P_i(0|p)}+k_B \int_{-\infty}^\infty dp \sum_{i=0}^{1}  P_i(p,t_m) \ln{P_i(t_m|p)}, \\ \label{know2}
&=-k_B \int_{-\infty}^\infty dp \sum_{i=0}^{1}  P_i(p,t_m) \ln{P_i(0|p)}+k_B \int_{-\infty}^\infty dp \sum_{i=0}^{1}  P_i(p,t_m) \ln{\frac{P_i(p,t_m)}{Q(p,t_m)}},\\ \label{known3}
&=k_B \int_{-\infty}^\infty dp \sum_{i=0}^{1}  P_i(p,t_m) \ln{\frac{P_i(p,t_m)}{Q(p,t_m)P_i(0|p)}}=k_B \int_{-\infty}^\infty dp \sum_{i=0}^{1}  P_i(p,t_m) \ln{\frac{P_i(p,t_m)}{\sum_i[P_i(p,t_m)]\int dpP_i(p,t_m)}}\geq0.
\end{align}
In Eq.\ \eqref{know2} we have used the fact that $P_i(0|p)$ is independent of $p$ and gives $P_0(0|p)=a=\int dp P_0(p,t_m)$ and $P_1(0|p)=b=\int dp P_1(p,t_m)$ which is equal to the marginal probabilities by tracing out the meter.  Eq.\ \eqref{known3} is the mutual information expression associated with the measurement process.
The second term in Eq.\ \eqref{know} reads
\begin{align}
\label{know4}
S(t_m)&=-k_B \int_{-\infty}^\infty dp \sum_{i=0}^{1}  P_i(p,t_m) \ln{\frac{P_i(p,t_m)}{Q(p,t_m)}} \\ \notag &=-k_B \int_{-\infty}^\infty dp \sqrt{\frac{1}{2 \pi  k_B T_M}}   a e^{-\frac{p^2}{2 k_B T_M}} \ln{\frac{a e^{-\frac{p^2}{2 k_B T_M}}}{a e^{-\frac{p^2}{2 k_B T_M}}+b e^{-\frac{(p+g t_m)^2}{2 k_B T_M}}}} \\ \notag &-k_B \int_{-\infty}^\infty dp \sqrt{\frac{1}{2 \pi  k_B T_M}}   b e^{-\frac{(p+g t_m)^2}{2 k_B T_M}} \ln{\frac{b e^{-\frac{(p+g t_m)^2}{2 k_B T_M}}}{a e^{-\frac{p^2}{2 k_B T_M}}+b e^{-\frac{(p+g t_m)^2}{2 k_B T_M}}}},
\end{align}
which in the limit $t_m\to \infty$ can be written as
\begin{align}
\label{know5}
S(t_m\to\infty)&=-k_B \int_{-\infty}^\infty dp \sqrt{\frac{1}{2 \pi  k_B T_M}}   a e^{-\frac{p^2}{2 k_B T_M}} \ln{\frac{a e^{-\frac{p^2}{2 k_B T_M}}}{a e^{-\frac{p^2}{2 k_B T_M}}}}-k_B \int_{-\infty}^\infty dp \sqrt{\frac{1}{2 \pi  k_B T_M}}   b e^{-\frac{(p+g t_m)^2}{2 k_B T_M}} \ln{\frac{b e^{-\frac{(p+g t_m)^2}{2 k_B T_M}}}{b e^{-\frac{(p+g t_m)^2}{2 k_B T_M}}}} \\ \notag &= 0,
\end{align}
so that the second term $S(t_m\to \infty)\to 0$ in Eq.\ \eqref{know} vanishes and $I(t_m\to\infty)=-k_B(a\ln a+b\ln b)$.

\section{Energy invest for measurement}
\label{energyshort}
As discussed in the main text the total Hamiltonian for the system, meter and their time-dependent coupling reads
\begin{align}
   \label{eq:totalHam}
   \hat{H}(t)=\hat{H}_S+\hat{H}_M+\hat{V}(t),
\end{align}
with 
\begin{align}
   \label{eq:coupling}
   \hat{V}(t)=\sum_i g_i(t)\ket{i}\bra{i},
\end{align}
while $g_i(t)\neq0$ during the measurement interval $0< t < t_m$.
The energy invest for the measurement is determined by the total change of energy during the measurement interval while $\hat{H}(0)=\hat{H}(t_m)=\hat{H}_S+\hat{H}_M$ and reads
\begin{align}
    \label{eq:meascost}
    W_{meas}(t_m)&\equiv \int_0^{t_M}dt \frac{d}{dt} \langle\hat{H}(t)\rangle = \int_0^{t_M}dt \frac{d}{dt} \mathrm{tr}[\hat{H}(t) \hat{\rho}(t)]=\mathrm{tr}[\hat{H}(t_m) \hat{\rho}(t_m)]-\mathrm{tr}[\hat{H}(0) \hat{\rho}(0)] \\ \notag
    &=\mathrm{tr}[\big[\hat{H}_S+\hat{H}_M \big] (\hat{\rho}(t_m)-\hat{\rho}(0))]=\mathrm{tr}[\hat{H}_M (\hat{\rho}(t_m)-\hat{\rho}(0))],
\end{align}
where $\hat{\rho}(t)=\hat{U}(t) \hat{\rho}(0)\hat{U}^\dagger(t)$ with $\hat{U}(t)=\exp{\{-i/\hbar \mathcal{T}\int_0^tdt'\hat{H}(t')\}}$ where $\mathcal{T}$ is the time-ordering operator while $\hat{\rho}(0)=\hat{\rho}_{S,in}\otimes \hat{\rho}_{M,in}$. Furthermore, the last equality in Eq.\ \eqref{eq:meascost} holds since $[\hat{H}_S,\hat{H(t)}]=0$.
Eq.\ \eqref{eq:meascost} can be alternatively written as 
\begin{align}
    \label{eq:meascost2}
    W_{meas}(t_m)&=\int_0^{t_M}dt \bigg[\mathrm{tr}[\dot{\hat{H}}(t) \hat{\rho}(t)]+\mathrm{tr}[\hat{H}(t) \dot{\hat{\rho}}(t)]\bigg]= \int_0^{t_M}dt \mathrm{tr}[\dot{\hat{V}}(t) \hat{\rho}(t)],
\end{align}
where we used $\dot{\hat{\rho}}(t)=-\frac{i}{h}[\hat{H}(t),\hat{\rho}(t)]$ and the cyclic permutation of the trace such that $\mathrm{tr}[\hat{H}(t) \dot{\hat{\rho}}(t)]=0$.

We can now apply Eq.\ \eqref{eq:meascost2} to a sudden switch on and off to/from a constant value as in our example in the main text where
$\hat{V}(t)=g\hat{x}\otimes\ket{1}\bra{1}\theta(t)\theta(t_m-t)=\hat{V}\theta(t)\theta(t_m-t)$. In this particular case Eq.\ \eqref{eq:meascost2} reads
\begin{align}
    \label{eq:meascost3}
    W_{meas}(t_m)&= \int_0^{t_M}dt \mathrm{tr}\bigg[\dot{\hat{V}}(t) \hat{\rho}(t)\bigg]=\bigg[ \mathrm{tr}[\hat{\rho}( 0) \hat{V}]-\mathrm{tr}[\hat{\rho}(t_m) \hat{V}] \bigg].
\end{align}
By using the Trotter splitting as in Eq.\ \eqref{P1} with $\Delta t =t_m/N$ we can further write Eq.\ \eqref{eq:meascost3} by
\begin{align}
\label{intE}
 W_{meas}( t_m)   =& -\bigg[ \mathrm{tr}[\hat{\rho}( t_m) \hat{V}]-\mathrm{tr}[\hat{\rho}(0) \hat{V}] \bigg] \\ \notag &=  
-b g \bra{D} \big(e^{i \hat{p}^2 \Delta t/2\hbar}e^{i\hat{V} \Delta t /\hbar}\big)^N  \hat{x} \big( e^{-i \hat{p}^2 \Delta t/2\hbar}e^{-i\hat{V} \Delta t /\hbar}\big)^N  \ket{D} 
\\ \notag 
&=  -b g \bigg (\frac{1}{2 \pi k_B T_M} \bigg )^{1/2}\bigg [\int_{-\infty}^\infty dp e^{-\frac{(p+g N \Delta t)^2}{4k_BT_M}}  \prod_{k=1}^N \bigg[ e^{i(p+g k \Delta t)^2\Delta t/2\hbar} \bigg] i \hbar \frac{d}{dp}  \\ \notag & \bigg\{\prod_{k=1}^N \bigg[ e^{-i(p+g k \Delta t)^2\Delta t/2\hbar} \bigg] e^{-\frac{(p+g N \Delta t)^2}{4k_BT_M}} \bigg\} \bigg] 
\\ \notag
&=  -b g \bigg (\frac{1}{2 \pi k_bT_M} \bigg )^{1/2}\bigg [\int_{-\infty}^\infty dp e^{-\frac{(p+g N \Delta t)^2}{2k_BT_M}}  \sum_{k=1}^N (p+g k \Delta t)\Delta t \bigg] 
\\ \notag
&=  -b g \bigg (\frac{1}{2\pi k_BT_M} \bigg )^{1/2}\bigg [\int_{-\infty}^\infty dp e^{-\frac{(p+g N \Delta t)^2}{2k_BT_M}}  \big( p N \Delta t+\sum_{k=1}^N g k \Delta t^2 \big) \bigg] 
\\ \notag
&=  -b g \bigg (\frac{1}{2\pi k_B T_M} \bigg )^{1/2}\bigg [\int_{-\infty}^\infty dp e^{-\frac{(p+g N \Delta t)^2}{2k_BT_M}}  \big( p t_m+ g N(N+1)(t_m/N)^2/2 \big) \bigg] 
\\ \notag
&=  -b g \bigg [  \big( -gt_m^2+ g t_m^2/2 + g t_m^2 /2N \big) \bigg] 
\\ \notag
& \underrel{\lim N \to \infty}{=}   \frac{bg^2 t_m^2}{2}.
\end{align}

We consider next the average energy change of the meter (change of kinetic energy of the free particle) $ \langle \Delta W_{M}(t_m) \rangle =\frac{1}{2}(\langle \hat{p}^2(t_m) \rangle-\langle \hat{p}^2(0) \rangle)$ after the entangling system-meter evolution of time $ t_m$ by using the Trotter splitting as in Eq.\ \eqref{P1} with $\Delta t =t_m/N$
\begin{align}
\label{P2}
 W_{M}(t_m) &=\frac{1}{2}\big[\mathrm{tr}[\hat{\rho}( t_m) \hat{p}^2]-\mathrm{tr}[\hat{\rho}(0) \hat{p}^2] \big]\\ \notag 
&= \frac{b}{2} \bigg[  \bra{D} \big(e^{i \hat{p}^2 \Delta t/2\hbar}e^{i\hat{V} \Delta t /\hbar}\big)^N   \hat{p}^2 \big( e^{-i \hat{p}^2 \Delta t/2\hbar}e^{-i\hat{V} \Delta t /\hbar}\big)^N  \ket{D}  \\ \notag &-  \bra{D} \hat{p}^2 \ket{D} \bigg]\\ \notag 
&=  \frac{b}{2}\bigg (\frac{1}{2 \pi k_B T_M} \bigg )^{1/2}\bigg [\int_{-\infty}^\infty dp e^{-\frac{(p+gN\Delta t)^2}{2 k_B T_M}} p^2 
-\int_{-\infty}^\infty dp e^{-\frac{p^2}{2 k_B T_M}} p^2\bigg]
\\ \notag 
&= \frac{b g^2 t_m^2}{2}.
\end{align}
Note that the $\langle \delta \hat{p}^2(t_m) \rangle = \langle \hat{p}^2 (t_m) \rangle - \langle p (t_m) \rangle ^2 \equiv 2 \langle \Delta W_M(t_m) \rangle $.

From Eq.\ \eqref{eq:meascost} we can identify $  W_{M}( t_m)\equiv W_{meas}( t_m) $. As consistency check we have shown that both Eqs.\ \eqref{P2} and \eqref{intE} leads to the same result.

\section{Work extraction by rethermalization}
\label{retherm}
Given the 2SS passive state after maximum work extraction under unitary transformation, we now discuss the possible work extraction from this 2SS passive state on the way to equilibrium with the thermal bath $T_S$.

Given the partition function for the 2SS $Z_S=1+\exp{(-\beta \Delta E)}$ of specific temperature $T=k_B^{-1}\beta^{-1}$, its heat capacity is given by
\begin{align}
\label{heatCap}
    C(T)=k_B \beta^2 \frac{d^2\ln{Z_S}}{d\beta^2}=  \frac{\Delta E^2}{k_B T ^2}\frac{e^{-\frac{\Delta E}{k_BT}}}{\big [1+e^{-\frac{\Delta E}{k_BT}}\big]^2}.
\end{align}

We recognize that any passive state of the 2SS is characterized by a positive temperature, $ T_p $. This depends on the outcome of the measurement and can be determine by
\begin{align}
\label{temperature}
    &T_p(p,t_m)=\frac{\Delta E}{k_B} \ln{\bigg[\frac{P_1(t_m|p)}{P_0(t_m|p)}\bigg]}^{-1} \Theta(P_1(t_m|p)-P_0(t_m|p))\\ \notag
    & + \frac{\Delta E}{k_B} \ln{\bigg[\frac{P_0(t_m|p)}{P_1(t_m|p)}\bigg]}^{-1} \Theta(P_0(t_m|p)-P_1(t_m|p)),
\end{align}
with $\Theta(x)$ being the heavy side function.
As the 2SS approaches thermal equilibrium with a bath at temperature $T_S $, it exchanges heat and its temperature changes. Recall that the maximum work $ W $ that can be generated from a heat transfer $ \mathcal{Q} $ between two baths at temperatures $ T_{\text{low}}$and $ T_{\text{high}} $ is given by the Carnot efficiency 
\begin{align}
W = \mathcal{Q} \left(1 - \frac{T_{\text{low}}}{T_{\text{high}}}\right)
\end{align}
We apply this principle to the heat bath at temperature $ T_S $ and the 2SS, acknowledging that the temperature of the 2SS changes as it absorbs and emits heat. Therefore, we must evaluate the work output incrementally, expressed as
\begin{align}
dW = d \mathcal{Q} \left(1 - \frac{T_{\text{low}}}{T_{\text{high}}}\right)
\end{align}
Here, we need to express $ d \mathcal{Q} $ in terms of the temperature change $ dT $. Specifically, we must distinguish between the scenarios where $ T_p > T_S $ and $ T_S > T_p $.

(i) $T_p > T_S$: In this scenario, $T_{low} = T_S$ and $T_{high} = T$, which changes from $T_p(p)$ to $T_S$. Using the relation $dQ = C(T) dT$, which represents the infinitesimal amount of heat flowing from the system into the bath, the maximum work that can be extracted is given by
\begin{align}
\label{thermalwork}
    W_{th}(p,t_m) = -\int_{T_p(p,t_m)}^{T_S} dT \, C(T) \left(1 - \frac{T_S}{T}\right).
\end{align}

(ii) $T_S > T_p$: In this scenario, $T_{low}=T$, which increases from $T_p$ to $T_S$, and $T_{high}=T_S$. Work can be extracted in this situation; however, the source of this work is the energy flowing out of the bath. When the bath transfers heat $dQ$, a portion of it, $dW = dQ(1 - T/T_S)$, is the maximum work that can be obtained. The remaining energy, $dQ_{sys} = dQ \frac{T}{T_S}$, is the heat entering the system, which raises the system's temperature by $dT = \frac{dQ_{sys}}{C(T)} = dQ \frac{T}{C(T) T_S}$. The maximum possible work extracted given measurement outcome $p$ is then given by
\begin{align}
\label{thermalwork2}
    W_{th}(p,t_m) = T_S \int_{T_p(p,t_m)}^{T_S} dT \frac{C(T)}{T} \left( 1 - \frac{T}{T_S} \right).
\end{align}

Surprisingly, in both scenarios, additional work is generated. In scenario (i), the source of energy is the system itself, whereas in scenario (ii), the energy comes from the bath. It is important to note that, because a Carnot process is involved, the time required to extract this work is infinite. In fact, both equations \eqref{thermalwork} and \eqref{thermalwork2} are the same and the average work extracted by rethermalization reads as

\begin{align}
\label{thermalwork3}
    W_{th} = \int_{-\infty}^\infty dp Q(p,t_m)\int_{T_p(p,t_m)}^{T_S} dT C(T) \left(\frac{T_S}{T} -1\right),
\end{align}

where the average is taken over all measurement outcomes $p$ whose probability density is given by $Q(p,t_m)$.

\section{Bound on work extraction}
\label{bound}

We can find an analytical solution for the integral \eqref{thermalwork2} by
\begin{align}
\label{integral2}
W_{th}(p,t_m) &=-\Delta E  \int_{-\Delta E/ k_B T_p}^{-\Delta E/ k_B T_S} dy \frac{e^y}{(1+e^y)^2}\bigg[ \frac{k_B T_S}{\Delta E}y +1 \bigg]  \\ \notag &
    = -\Delta E \bigg\{ \bigg[ \frac{ye^y}{1+e^y}  - \ln \bigg[ e^y+1\bigg]\bigg]_{-\Delta E/ k_B T_p}^{-\Delta E/ k_B T_S} \frac{k_BT_S}{\Delta E} +\bigg[ \frac{-1}{1+e^y}\bigg]_{-\Delta E/ k_B T_p}^{-\Delta E/ k_B T_S} \bigg \} 
    \\ \notag &
    = -\Delta E \bigg\{ \bigg[ \frac{\big( \frac{T_S}{T_p}-1\big) e^{-\Delta E/k_BT_p}}{1+e^{-\Delta E/k_B T_p}}   + \ln \bigg[ \frac{1+e^{-\Delta E/k_B T_p}}{1+e^{-\Delta E/k_B T_S}}\bigg]\frac{k_BT_S}{\Delta E}  \bigg \},
\end{align}
where we have used $y=-\Delta E/k_B T$ and $dy=\Delta E/k_B T^2 dT$.

Next we use the definition of the temperature $T_p(p,t_m)$ in Eq.\ \eqref{temperature} and evaluate Eq.\ \eqref{integral2} which then reads 
\begin{align}
\label{integral3}
&-\Delta E  \int_{-\Delta E/ k_B T_p(p,t_m)}^{-\Delta E/ k_B T_S} dy \frac{e^y}{(1+e^y)^2}\bigg[ \frac{k_B T_S}{\Delta E}y +1 \bigg] 
  \notag \\ & = k_BT_S \bigg\{ \bigg[ P_0(t_m|p)\ln P_0(t_m|p) + P_1(t_m|p)\ln P_1(t_m|p)   - P_0(t_m|p)\ln b - P_1(t_m|p)\ln a \bigg] \Theta(P_1(t_m|p)-P_0(t_m|p))  \\ \notag
     &+ \bigg[ P_0(t_m|p)\ln P_0(t_m|p) + P_1(t_m|p)\ln P_1(t_m|p)  - P_1(t_m|p)\ln b - P_0(t_m|p)\ln a \bigg] \Theta(P_0(t_m|p)-P_1(t_m|p))  \bigg \}.
\end{align}

The result of Eq.\ \eqref{integral3} can be used to evaluate the work extraction by thermalization in Eq.\ \eqref{thermalwork3} which then reads 
\begin{align}
\label{integral4}
&W_{th}= k_BT_S \int_{-\infty}^{\infty} dp Q(p,t_m) \bigg\{\bigg[ P_0(t_m|p)\ln P_0(t_m|p) + P_1(t_m|p)\ln P_1(t_m|p)   - P_0(t_m|p)\ln b - P_1(t_m|p)\ln a \bigg] \\ &\notag \times \Theta(P_1(t_m|p)-P_0(t_m|p))  + \bigg[ P_0(t_m|p)\ln P_0(t_m|p)+ P_1(t_m|p)\ln P_1(t_m|p)  - P_1(t_m|p)\ln b - P_0(t_m|p)\ln a \bigg]\\ \notag &\times \Theta(P_0(t_m|p)-P_1(t_m|p)) \bigg \}.
\end{align}
The result of Eq.\ \eqref{integral4} can be rearranged. By adding and subtracting $\bigg[ P_0(t_m|p)\ln P_0(t_m|p)+ P_1(t_m|p)\ln P_1(t_m|p)- P_1(t_m|p)\ln b - P_0(t_m|p)\ln a \bigg]\Theta(P_1(t_m|p)-P_0(t_m|p))$) and using the fact that $\Theta(P_1(t_m|p)-P_0(t_m|p))+\Theta(P_0(t_m|p)-P_1(t_m|p))=1$, Eq.\ \eqref{integral4} can be recast to
\begin{align}
\label{integral5}
&W_{th}= k_BT_S \int_{-\infty}^{\infty} dp Q(p,t_m)  \bigg\{ \bigg[P_1(t_m|p) - P_0(t_m|p) \bigg]\ln{\bigg(\frac{b}{a}\bigg)} \Theta(P_1(t_m|p)-P_0(t_m|p))   \\ \notag
     &+ \bigg[ P_0(t_m|p)\ln P_0(t_m|p)+ P_1(t_m|p)\ln P_1(t_m|p) - P_1(t_m|p)\ln b - P_0(t_m|p)\ln a \bigg] \bigg \}.
\end{align}

Using $\ln(b/a)=-\Delta E/k_BT_S$, exploiting the definition of the ergotropy Eq.\ (11) in the main text and realizing that the second term in Eq.\ \eqref{integral5} is the mutual information between system and meter as defined in Eq.\ (7) in the main text , we can recast Eq.\ \eqref{integral5} to
\begin{align}
\label{integral6}
W_{th}(t_m)= -W_{erg}(t_m)+T_S I(t_m),
\end{align}
or 
\begin{align}
\label{integral7}
W_{th}(t_m)+W_{erg}(t_m)= T_S I(t_m).
\end{align}

\end{document}